\documentclass{aa}
\usepackage[varg]{txfonts}
\usepackage{natbib}
\usepackage{graphicx}
\usepackage{times}
\bibpunct{(}{)}{;}{a}{}{,}
\newcommand{\br}{BR~CMi }
\newcommand{\bn}{BR~CMi}

\newcommand{\spefo}{{\tt SPEFO} }
\newcommand{\spefoe}{{\tt SPEFO}}
\newcommand{\phoebe}{{\tt PHOEBE} }
\newcommand{\phoebee}{{\tt PHOEBE}}
\newcommand{\fotel}{{\tt FOTEL} }

\newcommand{\korel}{{\tt KOREL} }
\newcommand{\korele}{{\tt KOREL}}

\newcommand{\tria}{\hbox{$\bigtriangleup$}}
\newcommand{\ubv}{\hbox{$U\!B{}V$}}
\newcommand{\bv}{\hbox{$B\!-\!V$}}
\newcommand{\ub}{\hbox{$U\!-\!B$}}
\newcommand{\uvby}{\hbox{$uvby$}}

\newcommand{\oc}{\hbox{$O\!-\!C$}}

\newcommand{\p}{$\pm$}

\newcommand{\m}{$^{\rm m}\!\!.$}

\newcommand{\ks}{km~s$^{-1}$}
\newcommand{\vsin}{$v$~sin~$i$ }

\newcommand{\tef}{$T_{\rm eff}$ }
\newcommand{\lgg}{{\rm log}~$g$ }
\newcommand{\ms}{M$_{\odot}$}
\newcommand{\rs}{R$_{\odot}$}

\newcommand{\ha}{H$\alpha$ }

\newcommand{\hae}{H$\alpha$}

\usepackage{color}

\begin{document}

   \title{Properties and nature of Be stars
\thanks{Based on new spectroscopic and photometric observations
from the following instruments:
Elodie spectrograph of the Haute Provence Observatory, France;
CCD coud\'e spectrograph of the Astronomical Institute AS~\v{C}R at
Ond\v{r}ejov, Czech Republic; CCD coud\'e spectrograph of the Dominion
Astrophysical Observatory, Canada;
HERMES spectrograph attached to the Mercator Telescope, operated on the
island of La Palma by the Flemish Community, at the Spanish Observatorio
del Roque de los Muchachos of the Instituto de Astrof\'\i sica de Canarias;
7-C photometer attached to the Mercator telescope, La Palma, \ubv\ photometers
at Hvar and Sutherland, and $H_{\rm p}$ photometry from the ESA Hipparcos
mission.}
%\fnmsep\thanks{Tables 3 and 5 are available only in electronic form
% at the CDS via anonymous ftp to cdarc.u-strasbg.fr (130.79.128.5)
% or via http://cdsweb.u-strasbg.fr/cgi-bin/qcat?J/A+A/}
}
\subtitle{30. Reliable physical properties of a semi-detached B9.5e+G8III
binary \br = HD~61273 compared to those of other well studied
semi-detached emission-line binaries}
\author{P.~Harmanec\inst{1}\and
P. Koubsk\'y\inst{2}\and
J.A.~Nemravov\'a\inst{1}\and
F.~Royer\inst{3}\and
D.~Briot\inst{3}\and
P.~North\inst{4}\and
P.~Lampens\inst{5}\and
Y.~Fr\'emat\inst{5}\and
S.~Yang\inst{6}\and
H.~Bo\v{z}i\'c\inst{7}\and
L.~Kotkov\'a\inst{2}\and
P.~\v{S}koda\inst{2}\and
M.~\v{S}lechta\inst{2}\and
D.~Kor\v{c}\'akov\'a\inst{1}\and
M.~Wolf\inst{1}\and
P.~Zasche\inst{1}
}
%  \thanks{Guest investigator, Dominion Astrophysical Observatory,
%          Victoria, BC, Canada}

   \offprints{P. Harmanec\,\\
               \email Petr.Harmanec@mff.cuni.cz}

  \institute{
   Astronomical Institute of the Charles University,
   Faculty of Mathematics and Physics,\hfill\break
   V~Hole\v{s}ovi\v{c}k\'ach~2, CZ-180 00 Praha~8 - Troja, Czech Republic
 \and
   Astronomical Institute, Academy of Sciences of the Czech Republic,
   CZ-251 65 Ond\v{r}ejov, Czech Republic
 \and
   GEPI/CNRS UMR 8111, Observatoire de Paris, Universit\'e Paris Denis Diderot,
   5 place Jules Janssen, F-92910~Meudon, France
 \and
  Laboratoire d'astrophysique, Ecole Polytechnique Fédérale de Lausanne (EPFL),
  Observatoire de Sauverny, CH-1290 Versoix, Switzerland
 \and
   Royal Observatory of Belgium, Ringlaan 3, B-1180 Brussel, Belgium
 \and
   Physics \& Astronomy Department, University of Victoria,
  PO Box 3055 STN CSC, Victoria, BC, V8W 3P6, Canada
 \and
  Hvar Observatory, Faculty of Geodesy, University of Zagreb,
  Ka\v{c}i\'ceva~26, 10000~Zagreb, Croatia
}
\date{Received \today}

  \abstract{
Reliable determination of the basic physical properties of hot emission-line
binaries with Roche-lobe filling secondaries is important for developing the theory of mass exchange in binaries. It is a very hard task, however,
which is complicated by the presence of circumstellar matter in these systems. So far,
only a small number of systems with accurate values of component masses, radii, and other
properties are known. Here, we report the first detailed study
of a new representative of this class of binaries, BR~CMi,
based on the analysis of radial velocities and multichannel
photometry from several observatories, and compare its
physical properties with those for other well-studied systems.
BR~CMi is an ellipsoidal variable seen under an intermediate orbital
inclination of $\sim51^\circ$, and it has an orbital period of 12\fd919059(15)
and a circular orbit. We used the disentangled component
spectra to estimate the effective temperatures 9500(200)~K
and 4655(50)~K by comparing them with model spectra.
They correspond to spectral types B9.5e and G8III. We also used
the disentangled spectra of both binary components as templates for
the 2-D cross-correlation to obtain accurate RVs and a reliable orbital
solution. Some evidence of a secular period increase at a rate of
($1.1\pm0.5$)~s per year was found. This, together with a very low mass ratio
of 0.06 and a normal mass and radius of the mass gaining component, indicates
that BR~CMi is in a slow phase of the mass exchange after the mass-ratio
reversal. It thus belongs to a still poorly populated subgroup of
Be stars for which the origin of Balmer emission lines is safely explained
as a consequence of mass transfer between the binary components.
}


\authorrunning{P. Harmanec et al.}
\titlerunning{Duplicity of the Be star \bn}
\maketitle

\section {Introduction}
In spite of the concentrated effort of several generations of astronomers, the very
nature of the Be phenomenon - the presence of the Balmer emission lines
in the spectra of some stars of spectral type B and their temporal variability
on several time scales - is still not well understood. The competing hypotheses
include (i) outflow of material from stellar photospheres, facilitated
either by rotational instability at the stellar equator, by stellar wind, or
by non-radial pulsations (or a combination of these effects), and (ii) several
versions of mechanisms facilitated by the duplicity of the objects in question.
It is true that the number of known binaries among Be stars is steadily
increasing,  but clear evidence that the Balmer emission is a consequence
of the binary nature of the Be star in question exists only for Be stars,
which have a mass-losing secondary that fills its Roche lobe. The number
of known systems of this type with reliably determined physical
properties is still rather small, and finding new representatives of
this subgroup of Be binaries is desirable, not only for understanding Be stars
but also from the point of view of the still developing
theory of mass transfer in semi-detached binaries (see, e.g.,
the recent studies by \citet{des2013} or \citet{davis2014}).
This paper deals with one such system.

\begin{table}
\begin{center}
\caption[]{Journal of RV data sets}\label{jourv}
\begin{tabular}{ccrccll}
\hline\hline\noalign{\smallskip}
Spg.&Time interval&No.   &Wavelength &Spectral  \\
 No.&             &of    & range     &res.\\
    &(HJD-2450000)&RVs   & (\AA)   \\
\noalign{\smallskip}\hline\noalign{\smallskip}
 1&0071.5--3430.4& 34&3850--6800&42000\\
 2&4000.6--5649.4& 56&6280--6730&12700\\
 3&4733.0--5994.8& 14&6150--6760&21700\\
 4&5278.4--5649.4& 19&8400--8900&17000\\
 5&5589.6--5912.7& 8&3800--8750&40000--90000\\
 3&6031.7--6383.8& 5&6310--6920&21700\\
\noalign{\smallskip}\hline\noalign{\smallskip}
\end{tabular}
\tablefoot{ Column ``Spectrograph No.": \ \
1...OHP 1.93-m reflector, Elodie echelle spg.;
2...Ond\v{r}ejov 2.0-m reflector, coud\'e grating spg., CCD detector,
    red spectra;
3...DAO 1.22-m reflector, coud\'e grating McKellar spg., CCD Site4 detector;
4...Ond\v{r}ejov 2.0-m reflector, coud\'e grating spg., CCD detector,
    infrared spectra;
5...Mercator 1.2-m reflector, HERMES echelle spg.
}
\end{center}
\end{table}

\begin{table}
\caption[]{Journal of available photometry for \bn.}\label{jouphot}
\begin{center}
\begin{tabular}{rcrccl}
\hline\hline\noalign{\smallskip}
Station&Time interval& No. of &Passbands&Ref.\\
       &(HJD$-$2400000)&obs.  & \\
\noalign{\smallskip}\hline\noalign{\smallskip}
61&47912.6--49061.9& 111&$V$  & 1 \\
37&52307.4--52384.4&  24&7-C   & 2 \\
93&52032.5--54836.8& 135&$V$  & 3 \\
11&55567.4--55579.5&  19&\ubv & 2\\
01&55574.5--56015.3& 103&\ubv & 2\\
\noalign{\smallskip}\hline
\end{tabular}\\
\tablefoot{Individual observing stations are distinguished by
the running numbers they have in the Prague / Zagreb photometric
archives --- see column ``Station":
01~\dots~Hvar 0.65-m, Cassegrain reflector, EMI9789QB tube;
11~\dots~SAAO 0.5-m reflector, \ubv\ photometry;
37\dots~Geneva 7-C system: observations secured with the Mercator reflector
at La~Palma;
61~\dots~Hipparcos all-sky $H_{\rm p}$ photometry transformed to Johnson $V$;
93~\dots~ASAS data archive \citep{pojm2002}.\\
 Column ``Ref.":
1...\cite{esa97};
2...this paper;
3...\cite{pojm2002}.
}
\end{center}
\end{table}

\br (HD~61273, BD+08$^\circ$1831, SAO~115739, IRAS~07361+0804, HIP~37232)
has not been studied much. Its light variability was discovered by
\cite{stet91} from his \uvby$\beta$ observations, but his individual
observations have never been published. The variability was rediscovered by
\citet{esa97} with the Hipparcos satellite, and a period of 6\fd46 was
reported by them.
\citet{kaz99} classified it as a $\alpha^2$~CVn variable and assigned it
a variable-star name \bn. \citet{paun2001} included this star into their
list of $\lambda$~Boo candidates but conclude that it is a normal star
of spectral type A6V.
In preliminary studies, \citet{briot2009} and \citet{roy2007} announced
that the object is actually a semi-detached Be dwarf and K0 giant
binary and an~ellipsoidal variable with an orbital period of 12\fd9190
and a mass ratio $M_{\rm K}/M_{\rm B}=0.14$. Hereafter, we call
the Be star component~1 and the cool giant component~2.
\citet{rens2009} included the object into their catalogue of Ap, HgMn, and Am
stars, giving it a spectral class A2, while \citet{dub2011}, who
were testing automatic classification procedures, classified it as
an~ellipsoidal variable with a period of 12\fd913.
We present the first detailed study of the system, based on
a rich collection of spectral and photometric observations.

%-------------------------------------------------------------------------------
\section{Observations and reductions}
{\sl Spectroscopy:}  The star was observed at the Haute Provence (OHP),
Ond\v{r}ejov (OND), Dominion Astrophysical (DAO), and La~Palma observatories.
Table~\ref{jourv} gives a journal of all RV observations. Details about
the spectra reductions and RV measurements are given in (online only)
Appendix~\ref{apa}.\\
{\sl Photometry:} We attempted to collect all available observations with known
dates of observations. Basic information about all data sets
can be found in Table~\ref{jouphot} and the details on the photometric
reductions and standardisation are in Appendix~\ref{apb}.
For convenience, we also publish all of our individual
observations together with their HJDs there
(in Tables~\ref{photg}, \ref{phothpv}, \ref{photv}, and \ref{photubv}).

\section{Finding reliable orbital elements}
One always has to be cautious when analysing binaries with clear signatures
of the presence of circumstellar matter in the system. The experience from
other such systems shows that the RV curve of the Roche-lobe filling component
is usually clean and defines its true orbit quite well, while the absorption
lines of the mass-gaining component can be affected by the circumstellar matter,
and their RV curves do not follow the true orbital motion
\citep[cf.][]{hs93,desmet2010}.

\begin{figure}
\centering
\includegraphics[width=9.0cm]{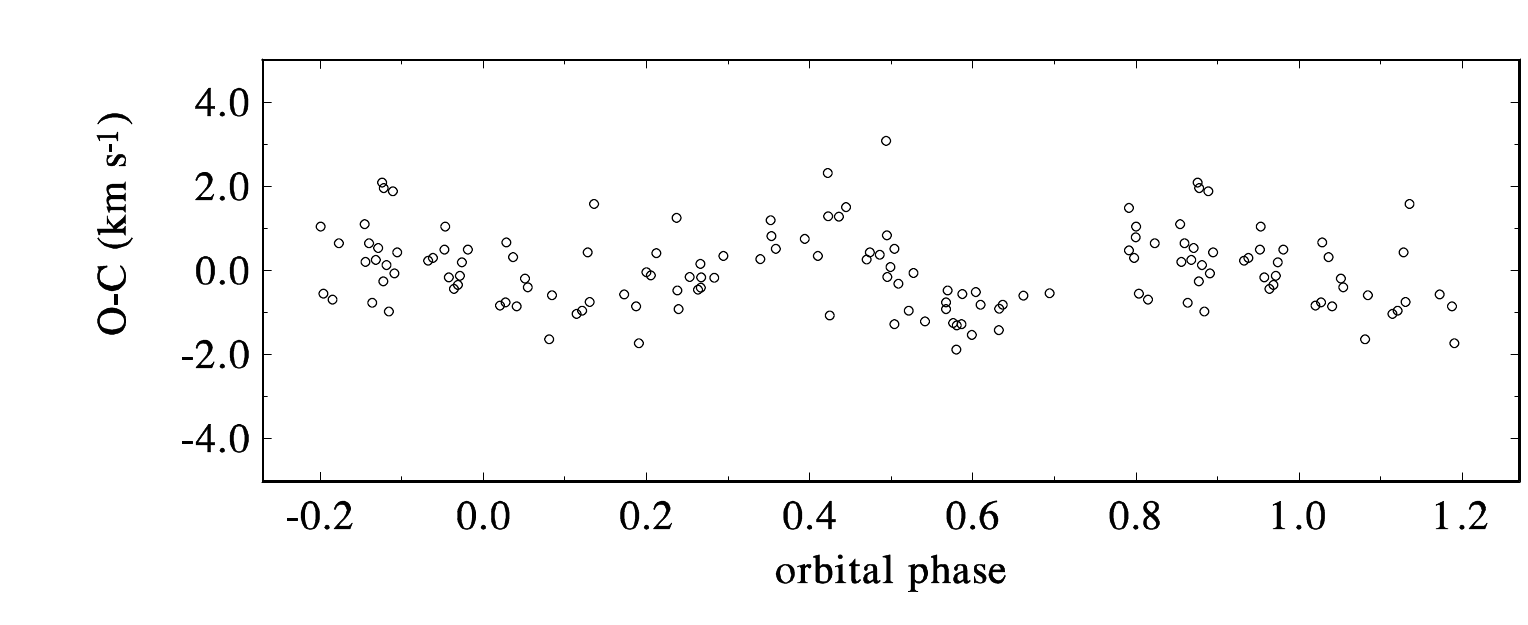}
\caption{\oc\ residuals from the circular-orbit
solution~2 for all RVs of component~2.}\label{rvspefo}
\end{figure}

To derive an improved ephemeris and an orbital solution for component~2,
we used all its RVs measured in \spefo (see Appendix~\ref{apa} for details)
and the program \fotel \citep{fotel1,fotel2}.
These solutions are summarised in Table~\ref{speforv}.
In solution~1, individual systemic ($\gamma$) velocities
were derived for each spectrograph. Since our correction of the RV zero point
via RV measurements of selected telluric lines clearly worked well
(the $\gamma$ velocities of all four instruments being nearly identical),
we assumed that all RVs are on the same system and derived solution~2 with
a joint systemic velocity. In this solution, we weighted the individual
datasets 1-4 by weights inversely proportional to the squares of rms errors
per one observation for each dataset from solution~1. As one can see
in Fig.~\ref{rvspefo}, the \oc\ residuals from the circular-orbit solution~2
define a low-amplitude double-wave phase curve. For that reason, we also
tentatively derived an eccentric-orbit solution. It indeed led to
an eccentricity of $e=0.0085\pm0.0012$, which is statistically significant
according to the test by \citet{lucy71}. However, the corresponding
longitude of periastron (referred to component~1) was
$\omega=93.3\pm8.5$~degrees, which is suspect. It is known that for tidally
distorted components of binaries, the difference between the optical and
mass centre of gravity leads inevitably to a deformation of the RV curve,
which manifests itself by  a false eccentricity with values of $\omega$
either 90$^\circ$ or 270$^\circ$ \citep{sterne41, bbin2003, eaton2008}.
This is what also occurs for \br and similar binaries, e.g. AU~Mon
\citep{desmet2010} or V393~Sco \citep{men2012b}.

Consequently, we adopted the circular-orbit solution~2 in Table~\ref{speforv}
as a basis for a working new ephemeris (\ref{efe1})

\begin{equation}
T_{\rm super.\,c.} = {\rm HJD}\,2454600.3806(30) +
                   12\fd919046(22) \cdot E\,.\label{efe1}
\end{equation}

\begin{table}
\caption[]{Trial \fotel orbital solutions based on RVs of component~2
measured in the program \spefoe.}
\label{speforv}
\begin{center}
\begin{tabular}{rcccccl}
\hline\hline\noalign{\smallskip}
Element           &  Solution 1            & Solution 2       \\
\noalign{\smallskip}\hline\noalign{\smallskip}
$P$ (d)           &12.919050\p0.000021     & 12.919046\p0.000022 \\
$T_{\rm super.\, c.}$&54600.3786\p0.0029&54600.3806\p0.0030\\
$e$               & 0.0 assumed            & 0.0 assumed      \\
$K_2$ (\ks)       &89.2086\p0.0068         & 89.1787\p0.0070  \\
$\gamma_1$ (\ks)    &$-$14.04\p0.12        &     --           \\
$\gamma_2$ (\ks)    &$-$14.12\p0.15        &     --           \\
$\gamma_3$ (\ks)    &$-$14.02\p0.13        &     --           \\
$\gamma_5$ (\ks)    &$-$14.49\p0.28        &     --           \\
$\gamma_{\rm joint}$ (\ks)    & --         & $-$14.088\p0.088 \\
rms (\ks)         &0.905                   & 0.781            \\
No. of RVs        & 117                    & 117              \\
\noalign{\smallskip}\hline\noalign{\smallskip}
\end{tabular}
\tablefoot{All epochs are in RJD = HJD-2400000;
rms is the rms per 1 observation of unit weight. The systemic velocities
$\gamma$ of individual spectrographs are identified by their numbers
used in Table~\ref{jourv}.}
\end{center}
\end{table}

\begin{figure}
\centering
\resizebox{\hsize}{!}{\includegraphics{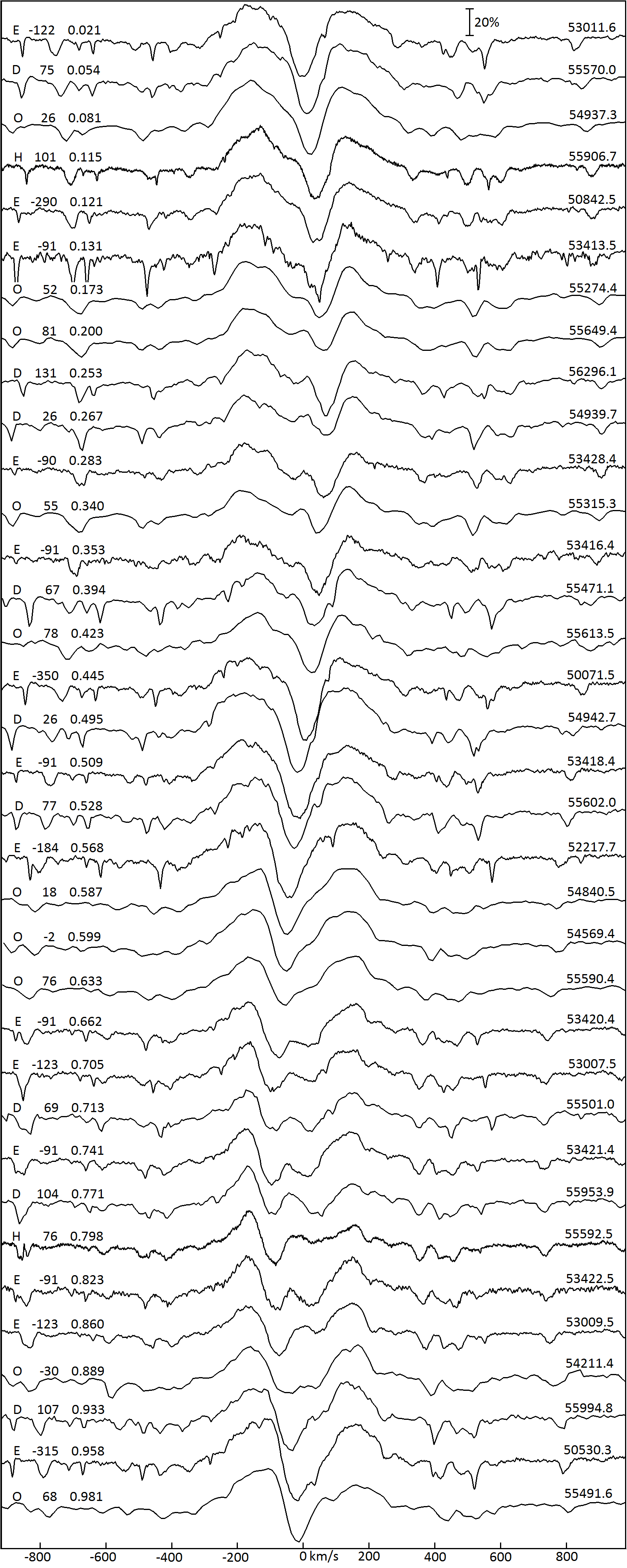}}
\caption{Selected \ha line profiles of \br stacked with increasing
orbital phase. Profiles are identified by cycle numbers and orbital phases
with respect to ephemeris~(\ref{efe1}) and by RJDs=HJDs$-2400000.0$\,.
Symbols E,O,D, and H stand for Elodie, Ond\v{r}ejov, DAO, and Hermes
spectrographs.}\label{ha}
\end{figure}

 Next we measured and analysed the \ha profile. Selected \ha profiles
stacked with increasing orbital phase are shown in Fig.~\ref{ha}.
For binary Be stars with strong \ha emission, this is the best measure of
the true orbital motion of the Be component  -- see \citet{bozic95,zarf26}
and \citet{peters2013}. This is not so for \bn, since its
\ha emission rises to only 20~\% above the continuum level, and the measured
RV curve of the \ha emission wings is poorly defined and has a phase shift
with respect to expected orbital motion of component~1.
In Fig.~\ref{ha} one can also note that the width of both peaks of the
double \ha emission varies systematically with the orbital phase, being
largest shortly after both conjunctions. Figure~\ref{ha} also shows
the secular stability of the strength of \ha emission, with the observed changes
mainly due to RV shifts of two absorption components, the stronger one
associated with component~2 and a weaker one associated with material
in the neighbourhood of component~1. It exhibits a RV curve, which is not
in exact antiphase to that of component~2.

In another attempt we used the program \korele\footnote{The freely distributed version from
Dec. 2004} developed by \citet{korel1, korel3} for spectra
disentangling. Since the expected semi-amplitude of the orbital RV curve
of component~1 is low, we decided to only analyse high-resolution
Elodie and Hermes spectra from the blue spectral region between
4397 and 4608~\AA. This region contains a number of \ion{He}{i} and metallic
lines but no Balmer line (to avoid contamination by the circumstellar
matter). It turned out that because of relatively broad and not very
numerous line profiles of component~1 and a low semi-amplitude of its RV curve,
comparably good solutions could be obtained for a wide range of mass ratios.
We note, however, that for the same reasons  the disentangling of
individual spectra is robust and can be trusted.

\begin{figure*}
\centering
\resizebox{\hsize}{!}{\includegraphics{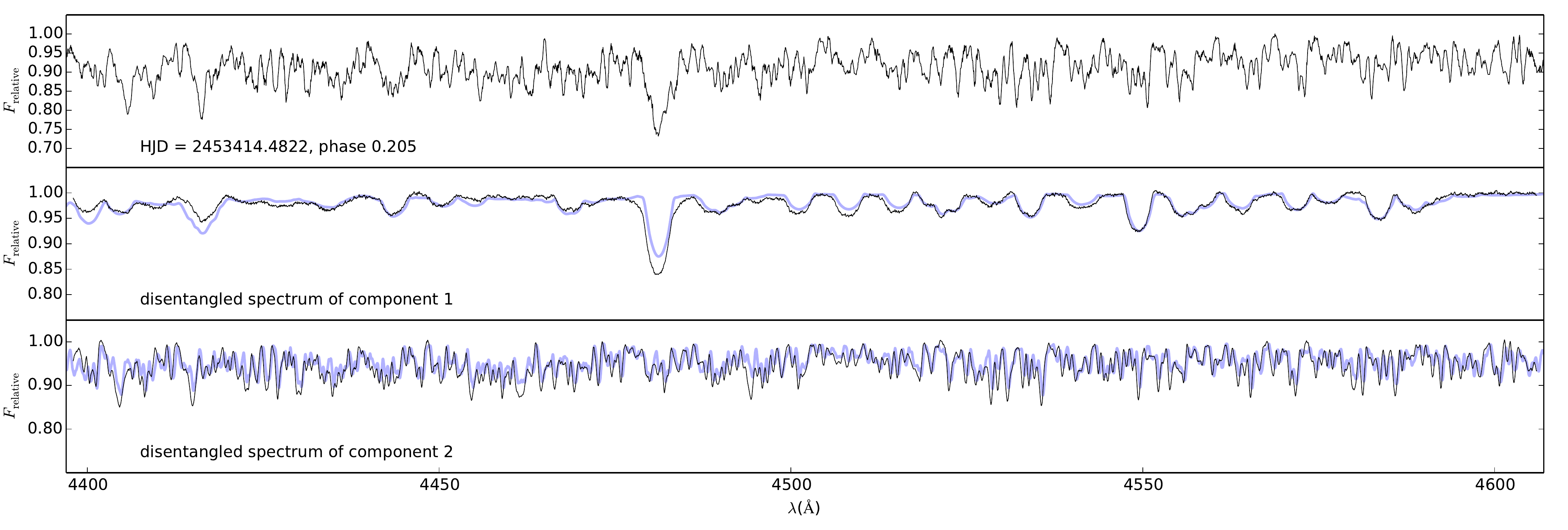}}
\caption{One observed Elodie spectrum and the disentangled spectra
of components~1 and~2 in the wavelength interval 4397 - 4608~\AA.
The disentangled spectra were renormalised to their own continua
and are compared to a model fit (see Sect.~\ref{fit}).}\label{todcor}
\end{figure*}

We then decided to try a cross-correlation in two dimensions, suggested (and
realized as the program {\tt TODCOR}) by \cite{todcor1}. In this technique,
the individual observed spectra are compared to the two template spectra,
which should closely correspond to the spectra of the two binary components,
both in the spectral type and the projected rotational broadening. The RVs
for both binary components are derived for each observed spectrum. A
similar program to {\tt TODCOR} was written by one of us (YF) under the name
{\tt asTODCOR} and has already been successfully applied to the case of AU~Mon
\citep{desmet2010}. In practice, the Fourier transform of the observed and
template spectra is computed via the FFT technique \citep{press93}.
To meet the requirements of the FFT algorithm \citep[see
e.g.][]{david2014}, the spectra were rebinned to a smaller, constant,
$\log \lambda$ bin size and apodised at both edges by means of a cosine-bell
function. \citet{david95} showed that the combination of an even and uneven
number of points correctly chosen around the CCF summit allows reducing
the impact of discretisation on the measurement of its location,
hence on the RV determinations. This approach was therefore also
implemented in {\tt asTODCOR} by combining the results of a parabola
fit through three and four points.

\begin{figure}
\centering
\includegraphics[width=9.0cm,angle=-180]{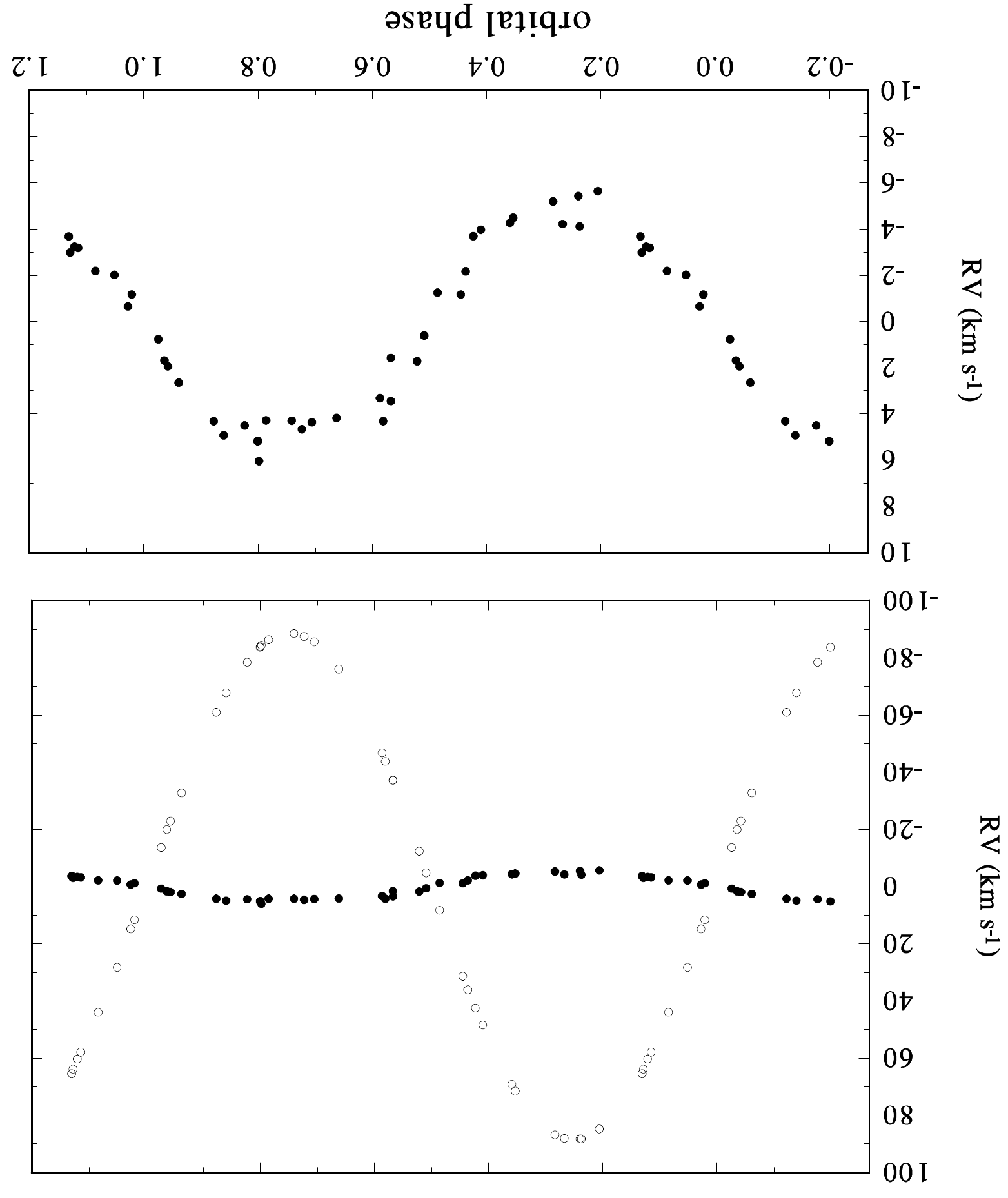}
%\resizebox{\hsize}{!}{\includegraphics{tod1.eps}}
\caption{{\sl Top:} Orbital RV curve of both components of \br based
on RVs derived via 2-D cross-correlation. {\sl Bottom:} Zoomed RV curve
of component~1 from the upper panel showing how accurately the RVs were
derived via the 2-D cross-correlation.}\label{rvtod}
\end{figure}

It is usual practice to use suitable synthetic and properly rotationally
broadened spectra as the templates. In this case, we used the spectra
of both components disentangled by \korel as the templates,
adopting a~preliminary estimated luminosity ratio
$L_2/L_1=0.25$. The wavelength interval 4397 -- 4608~\AA\ was used
(see Fig.~\ref{todcor}), and we verified
that the disentangled spectrum of component~1 had the same systemic velocity
as component~2. The procedure worked remarkably well and returned well-defined
RV curves for both binary components.
This is seen in Fig.~\ref{rvtod}, which is the phase plot of
these curves for ephemeris (\ref{efe1}). We publish all individual \spefo
RVs for component~2 and {\tt asTODCOR} RVs for both components, together
with their HJDs in Tables~\ref{rvsp} and \ref{rvtd} in Appendix~\ref{apa}.

In Table~\ref{soltod} we provide the circular-orbit solutions based on
RVs resulting from 2-D cross-correlation.
Solution~3 is based on RVs of component~1 and solution~4 on RVs
of component~2 alone. Their comparison shows that the epochs of superior
conjunction of component~1 from both solutions agree with each other
within the limits of their respective errors. They also agree with the epoch
derived from a more numerous set of RVs from the red spectral region
measured in \spefo (see solution~2 of Table~\ref{speforv}). Since the 2-D
cross-correlation puts no constraint on the derived RVs, this seems to
prove that the RVs of component~1, found to be in exact antiphase to
those of component~2, indeed reflect the true orbital motion of component~1.
In addition, both solutions have identical RV zero points within its rms
error of 0.1~\ks, confirming that using disentangled spectra has not
introduced any biases.
We therefore derived another solution~5, which is based on RVs for
both binary components. It defines the $M_{1,2}\sin^3i$ values and the
mass ratio $M_2/M_1=0.0593$.

\begin{table}
\caption[]{\fotel circular-orbit solutions individually for components~1 and~2
and for both stars together, based on RVs obtained via a 2-D
cross-correlation. The orbital period was fixed at 12\fd919046
in all cases.}\label{soltod}
\begin{center}
\begin{tabular}{rrrrccl}
\hline\hline\noalign{\smallskip}
Element         &  Solution 3 & Solution 4 & Solution 5   \\
\noalign{\smallskip}\hline\noalign{\smallskip}
$T_{\rm sup.\, conj.}$&  0.427   &  0.3807    &  0.3808   \\
                      &\p0.059   &\p0.0032    &\p0.0033   \\
$K_1$ (\ks)           &5.26\p0.15&    --      &5.26\p0.14 \\
$K_2$ (\ks)           &    --    & 88.71\p0.14&88.71\p0.15\\
rms (\ks)             &0.694     & 0.630      & 0.666     \\
No. of RVs            & 41       & 41         & 41/41     \\
\noalign{\smallskip}\hline\noalign{\smallskip}
\end{tabular}
\tablefoot{All epochs are in HJD-2454600;
rms is the rms of 1~observation of unit weight.}
\end{center}
\end{table}

\section{An estimate of basic physical
elements of the binary components and the whole system}

\begin{table}
\centering
\caption{Average results of the final best fit of the interpolated grid
of synthetic spectra to the observed ones in the six considered spectral
regions. The majority of the physical properties affecting the line profiles
were kept fixed from the final combined orbital and light-curve solution.
\tef of component~2 is based on the fit in the IR region alone -- see the text
for details.}
\begin{tabular}{lr@{$\pm$}l|r@{$\pm$}l}
\hline\hline\noalign{\smallskip}
Quantity& \multicolumn{2}{c}{Component 1} & \multicolumn{2}{c}{Component 2}\\
\noalign{\smallskip}\hline\noalign{\smallskip}
$T_{\rm eff}$\,(K)&9514&171  &4655&50\\
% $\log (g_{\rm [cgs]})$&3.738&0.014  &3.706&0.009\\
$v\sin i$\,(\ks)&130.98&3.27  &20.11&2.71\\
% $L_{\rm R}$&0.5712&0.0021    &0.4288&0.0021\\
% $RV$\,(\ks)&$-20.2$&2.6  &$-19.72$&0.36\\
% $Z$\,(Z$_{\odot}$)&\multicolumn{2}{c}{1.00} &\multicolumn{2}{c}{1.00}\\
\noalign{\smallskip}\hline
\end{tabular}
\label{spfit}
\end{table}

\begin{table}
\centering
\caption{Combined radial-velocity curve and light curve solution.
The optimised parameters are given with only their formal errors.
}
\label{phoebe}
\begin{tabular}{lr@{$\pm$}lr@{$\pm$}l}
\hline
Element& \multicolumn{4}{c}{Orbital properties}\\
\hline\hline
$P$\,(d)& \multicolumn{4}{c}{12.919059$\pm\substack{0.000015\\0.000015}$}\\
$T_{\rm min}$\,(RJD)& \multicolumn{4}{c}{54600.38186$\pm\substack{0.00206\\0.00206}$}\\
$M_2/M_1$\,& \multicolumn{4}{c}{0.0593$^*$}\\
$i$\,(deg)& \multicolumn{4}{c}{50.85$\pm\substack{5.33\\2.70}$}\\
$a$\,(\rs)& \multicolumn{4}{c}{31.21$\pm\substack{1.28\\2.08}$}\\
\hline
\hline
& \multicolumn{4}{c}{Component properties}\\
& \multicolumn{2}{c}{Component 1}& \multicolumn{2}{c}{Component 2}\\
\hline
\tef\,(K)& \multicolumn{2}{c}{9685$^*$}& \multicolumn{2}{c}{4655$^*$}\\
$\Omega$& 277.2&$\substack{5.0\\26.8}$& \multicolumn{2}{c}{22.82$^*$}\\
$\log g_{\rm \left[cgs\right]}$& 4.228&$\substack{0.031\\0.118}$& 2.095&$\substack{0.018\\0.033}$\\
$M$\,(\ms)& 2.31&$\substack{0.29\\0.43}$& 0.137&$\substack{0.017\\0.026}$\\
$R$\,(\rs)& 1.93&$\substack{0.13\\0.00}$& 5.49&$\substack{0.22\\ 0.37}$\\
$F$& \multicolumn{2}{c}{22.40$^{**}$}& \multicolumn{2}{c}{1.00$^{**}$}\\
$M_{\rm bol}$&1.080&$\substack{0.101\\0.066}$&1.991&$\substack{0.152\\0.085}$ \\
$L_{\rm Johnson-U}$ &0.953&$\substack{0.008\\0.009}$ &0.047&$\substack{0.009\\0.008}$\\
$L_{\rm Johnson-B}$ &0.890&$\substack{0.018\\0.014}$ &0.110&$\substack{0.014\\0.018}$\\
$L_{\rm Johnson-V}$ &0.740&$\substack{0.035\\0.022}$ &0.260&$\substack{0.022\\0.035}$\\
$L_{\rm Geneva-U}$  &0.994&$\substack{0.001\\0.001}$ &0.006&$\substack{0.001\\0.001}$\\
$L_{\rm Geneva-B1}$ &0.984&$\substack{0.003\\0.003}$ &0.016&$\substack{0.003\\0.003}$\\
$L_{\rm Geneva-B}$  &0.963&$\substack{0.007\\0.006}$ &0.037&$\substack{0.006\\0.007}$\\
$L_{\rm Geneva-B2}$ &0.936&$\substack{0.011\\0.009}$ &0.064&$\substack{0.009\\0.011}$\\
$L_{\rm Geneva-V1}$ &0.816&$\substack{0.027\\0.018}$ &0.184&$\substack{0.018\\0.027}$\\
$L_{\rm Geneva-V}$  &0.800&$\substack{0.029\\0.019}$ &0.200&$\substack{0.019\\0.029}$\\
$L_{\rm Geneva-G}$  &0.739&$\substack{0.036\\0.021}$ &0.261&$\substack{0.021\\0.036}$\\
$L_{\rm Hp}$        &0.804&$\substack{0.029\\0.022}$ &0.196&$\substack{0.021\\0.029}$\\
\hline
\end{tabular}
\tablefoot{\\
$^*$ Parameters that were fixed during the optimisation.\\
$^{**}$ The synchronicity parameter $F$ of component~1 was kept fixed
during each iteration, but it was recalculated for new radius and inclination
after each iteration, using \vsin from Table~\ref{spfit}. A spin-orbit
synchronization was assumed for the Roche lobe-filling component~2.
}
\end{table}

Since \br is not an eclipsing but an ellipsoidal binary, its
light curves can almost equally well be fitted for a wide range
of possible orbital inclinations. For that reason, we proceeded
in an iterative way, trying to describe all system properties in
a mutually consistent way and taking all limitations we
have at our disposal into account. In particular, we used the disentangled spectra of
both binary components in several spectral regions to obtain
good estimates of the effective temperatures of both stars via
a comparison of disentangled and synthetic line profiles. These were then
kept fixed in the combined light-curve and orbital solutions for a given
inclination. Since the masses and radii thus obtained define \lgg quite
accurately, we kept them fixed in the determination of $T_{\rm eff}$.

\subsection{Comparing the disentangled and
interpolated model spectra}\label{fit}
We used a program of J.A.~Nemravov\'a \citep[see][]{anita2014}
that interpolates in a~precalculated
grid of synthetic stellar spectra sampled in the effective temperature,
gravitational acceleration, and metallicity.
The synthetic spectra are compared to the disentangled ones for two
or more components of a multiple system (normalised to the joint continuum
of all bodies) and the initial parameters are optimised by minimisation
of $\chi^2$ until the best match is achieved.

\begin{figure}
\centering
\includegraphics[width=10cm]{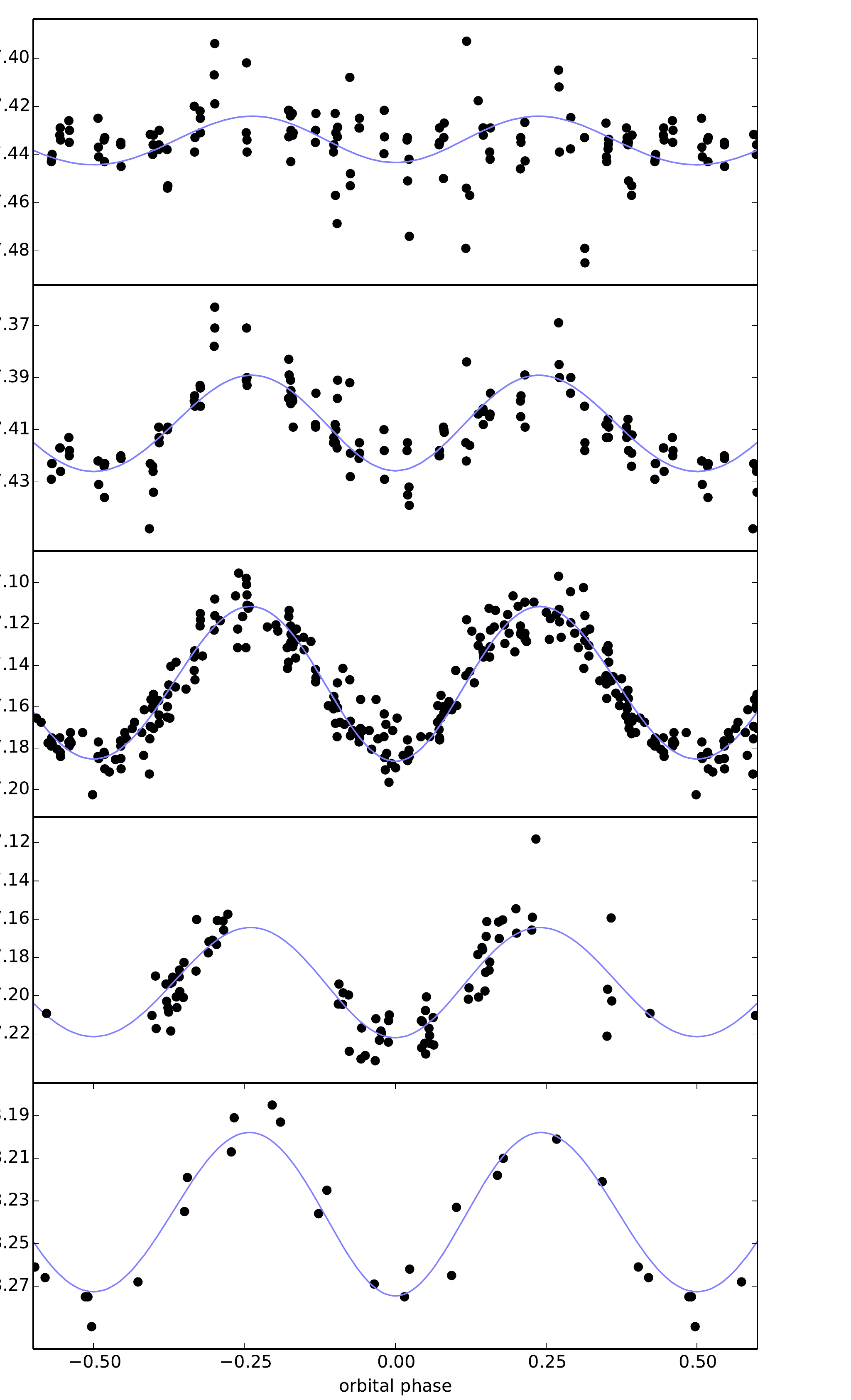}
\caption{Comparison of the photometric observations (black points) with
the \phoebe model light curve corresponding to the $\chi^2$ minimum
for several photometric passbands (solid/blue line). The fits for
the remaining passbands look similar. Phases from ephemeris~\ref{efe2}
are used. Note the colour dependence of the
amplitude of photometric changes on the wavelength.}
\label{lcfit}
\end{figure}

Since the fractional luminosities within a considered spectral region are
treated as independent of wavelength, we fitted the following six
disentangled spectral bands separately:
4002-4039, 4142-4191, 4381-4435, 4446-4561, 4910-4960, and 8400-8880\,\r{A}.
The interpolation was carried out in two different grids of spectra:
AMBRE grid computed by \citet{delaverny2012} was used for component~2,
and the Pollux database computed by \citet{palacios2010} was used for
the hotter component~1.

It would not be wise to derive all physical parameters
via free convergence in all considered spectral regions. Since the
combined light and orbital solution can provide very tight
values of \lgg and systemic velocity $\gamma$, we kept these values fixed
in each superiteration between model spectra and combined solution.
The infrared region 8400-8880\,\r{A} is the most appropriate region for
the \tef determination of component~2, since its fractional luminosity is
much higher there ($\sim0.5$ compared to less than 0.2 in the blue regions)
and its spectrum is rich there.

Having the effective temperature of component~2 determined with an accuracy
of $\sim 50$~K, we kept it fixed in all five blue spectral regions, where the
fractional luminosity of component~2 is low.
The uncertainty of the fitted parameters has two
components. The first is the statistical one connected with the photon
noise and represents a minor contribution since the disentangled profiles
have very high S/N. The second is influenced by the approach to the
minimisation and can be estimated by
repeating the fitting procedure and the need for (re)normalisation
of the disentangled profiles\footnote{Since the disentangling in
\korel is carried out in Fourier space, it happens in almost every
practical application that the resulting disentangled spectra after
inverse Fourier transformation have slightly warped continua and need
some re-normalisation.}. The former source of systematic error
is not taken into account, because our program does not
normalise the disentangled spectra. Therefore, we estimated the
uncertainties of the optimised parameters as a mean of the
best fit values obtained for each modelled spectral band,
which should (at least to some degree) account for the
need for (re)normalisation of the disentangled profile. The only
exception is the uncertainty of the effective temperature of component~2,
which was estimated via a simple Monte Carlo simulation. This is justified
since the disentangled continua of component~2 were very close to
a straight line in all regions. The results of the final minimisation averaged
over all considered spectral regions are summarised in Table~\ref{spfit}.

\begin{table*}
\caption{List of some basic properties of well-observed semi-detached
emission-line binaries. In all cases, component 1 is the gainer and
component~2 is the Roche-lobe-filling mass losing star. The mass
ratio is always the mass of the mass-losing star over the mass of the
gainer.}\label{emisbin}
\begin{tabular}{rrrrccccccrc}
\hline\hline\noalign{\smallskip}
  Binary & HD &Sp1+Sp2&$P_{\rm orb}$&$\dot P$&Cyclic&$M_1+M_2$&$M_2/M_1$
&$r_1/r_1^{\rm Roche}$&$i$&Principal\\
         &    &       &     (d)     & (s/yr)&changes(d)&(\ms)& & &($^\circ$)&references\\
\noalign{\smallskip}\hline\noalign{\smallskip}
  RX Cas&  --  &B5:+K1III &32.31&+19.86 & 516.1  &5.8+1.8&0.30   &0.06&80&1\\
  SX Cas&232121&B5:+K3III &36.56& -4.80 &  ?     &5.1+1.5&0.30   &0.07&79&2\\
  AU Mon& 50846&B5+G2III  &11.11&  0.0  & 416.9  &7.0+1.2&0.17   &0.24&79&3\\
  BR CMi& 61273&B9.5+G8III&12.92&  0.0  &   no   &2.31+0.14&0.06 &0.08&51&4\\
  UX Mon& 65607&B7:+K0III &5.904&-0.260 &  yes   &3.38+3.90&1.15 &0.36&81&5\\
  TT Hya& 97528&B9.5+G5III&6.953&  0.0: &  ?     &2.77+0.63&0.23 &0.17&83&6\\
V393 Sco&161741&B3+A8III  &7.713&$<0.5$ & 253.0  &7.8+2.0&0.25   &0.25&80&7\\
  CX Dra&174237&B3+F5III  &6.696&  0.0  &  yes   &7.3+1.7:&0.23  &0.25&53&8\\
$\beta$ Lyr&174638&B0:+B6-8II&12.94&+18.93& 282.4&13.1+2.96&0.23 &  ? &86&9\\
V448 Cyg&190967&O9.5V+B1Ib&6.520&  0.0  &   no   &24.7+13.7&0.56 &0.37&88&10\\
V360 Lac&216200&B4:+F9III &10.09& +5.5? & 322.2  &7.45+1.21&0.16 &0.48&65&11\\
\hline\noalign{\smallskip}
\end{tabular}
\tablefoot{
1...\citet{ander89},
2...\citet{ander88},
3...\citet{desmet2010, atw2012},
4...this paper,
5...\citet{olson2009, sudar2011},
6...\citet{miller2007},
7...\citet{men2012a,men2012b},
8...\citet{kou98,rich2000},
9...\citet{hec96,bis2000,hec2002,ak2007},
10...\citet{djuras2009},
11...\citet{zarf17,zarf24}\,.
}
\end{table*}

\subsection{A combined orbital and light-curve solution}
We used the program \phoebe \citep[version 1.0 with a few custom
passbands added;][]{prsa2005,prsa2006},
all photometric data sets available to us, RVs of both components from the 2-D
cross-correlation and also RVs of component~2 measured in SPEFO because they
give almost the same RV curve as the RVs from 2-D cross-correlation but
represent a much more numerous data set. For all Elodie and Hermes spectra,
for which both sets of RVs of component~2 exist, we set their weights to 0.5
in both subsets.  We assumed that component~2 fills
the Roche lobe, fixed the mass ratio $M_2/M_1=0.0593$, which follows from
the orbital solution for the cross-correlated RVs from spectra with high
S/N (see solution~5 of Table~\ref{soltod}).

An unsuccessful attempt to converge \tef of component~1 was carried out.
Changes of the \tef did not lead to any improvement in the cost function value,
and the program merely adjusted the radii of both components to
adapt the model to the new temperature. We thereforee decided to
fix the effective temperatures of both components obtained from
the fit of synthetic spectra to the disentangled ones.
We adopted \tef$\,=\,4655$~K for component~2. The effective
temperature of component 1 is rather uncertain. We therefore computed
three sets of trial solutions for three different \tef of the
component~1 corresponding to the estimated 1$\sigma$ bars given in
Table~\ref{spfit}: \tef$\,\epsilon\,\lbrace9343, 9514, 9685\rbrace$~K.
 The solution that has the least $\chi^2$ of all trial solutions
is listed in Table~\ref{phoebe}. Error bars presented in Table~\ref{phoebe}
were estimated as the difference between the maximal (minimal) result
of each parameter, which did not have the $\chi^2$ greater than 1\% of the $\chi^2$ of the
best solution, and the best solution. The bolometric albedos were estimated
for the corresponding effective temperatures from Fig.~7 of \citet{claret2001}
and the coefficients of gravity brightening from Fig.~7 of \citet{claret98}.
Logarithmic limb-darkening coefficients were automatically interpolated
after each iteration in \phoebe from a pre-calculated grid of model
atmospheres.

A test of the robustness of the final solution presented in Table~\ref{phoebe}
was carried out using the scripter environment of \phoebee. We fitted
the photometric observations and RVs from the 2-D cross-correlation
starting the convergence for 29 different initial orbital
inclinations $i$ ranging from 30 to 86~degrees. During each run,
30 iterations were computed, although the solution usually converged in
less than ten iterations. The rotation parameter $F$ (a ratio of spin to
orbital revolution) of component~1 was not converged in \phoebe, but it was
optimised after each iteration using the new values of the component~1 radius
and of the orbital inclination, while keeping the value of \vsin fixed
from Table~\ref{spfit}.
A comparison of the optimal \phoebe model light curves with the observed ones
is in Fig.~\ref{lcfit}.

The free convergence of the orbital
period and epoch, now based not only on RVs but also on photometry, led
to an improved linear ephemeris (\ref{efe2})
\begin{equation}
T_{\rm super.\,c.} = {\rm HJD}\,2454600.38187(42) +
                   12\fd9190594(7) \cdot E\,.\label{efe2}
\end{equation}
To check on a secular constancy of the orbital period, we split the RVs
and photometry into two subsets: HJD~2448705-53650 containing 460 data,
and HJD~2454000-56295 containing 505 data. This particular choice ensured
that both subsets contain a representative mixture of RVs and photometry.
Local \phoebe solutions for these two subsets led to the following
linear ephemerides (\ref{efe3}) and (\ref{efe4})
\begin{eqnarray}
T_{\rm super.\,c.} &=& {\rm HJD}\,2450724.6695(47) +
                   12\fd919018(30) \cdot E,\label{efe3}\\
T_{\rm super.\,c.} &=& {\rm HJD}\,2454600.37787(30) +
                   12\fd919165(45) \cdot E.\label{efe4}
\end{eqnarray}
\noindent They imply a secular increase in the orbital
period for $3.8\cdot 10^{-8}$ days per day or 1.2 s\,yr$^{-1}$. We therefore
ran another \phoebe solution on the complete data set, this time
also allowing for the convergence of the secular period change. It provided
a marginal support for the reality of a secular period increase
of $(3.5\pm1.6)\cdot10^{-8}$~d\,d$^{-1}$, the corresponding quadratic
ephemeris (\ref{efe5}) being
\begin{eqnarray}
T_{\rm super.\,c.} &=& {\rm HJD}\,2454600.3800(22)\nonumber\\
&+& 12\fd919107(26) \cdot E + 2.31\cdot10^{-7}\cdot E^2\,.\label{efe5}
\end{eqnarray}

The original Hipparcos parallax \citep{esa97} is 0\farcs00670$\pm$0\farcs00088.
In \citet{leeuw2007a, leeuw2007b} an improved value of the
Hipparcos parallax of \bn, 0\farcs00579$\pm$0\farcs00061 is given.
These values imply a distance $132 - 172$~pc for the original,
and $156 - 193$~pc for the improved value. A standard dereddening of
\ubv\ magnitudes of component~1 using the fractional luminosities of
Table~\ref{phoebe} gives
$V_0=7$\m08, (\bv)$_0=-0$\m03, and (\ub)$_0=-0$\m11\,.
These colours correspond to a normal B9.5 star, in agreement with the
mean \tef=9685~K adopted as the best from the combined model fit
of Table~\ref{phoebe} for component~1.  Consulting the tabulation of
normal stellar masses and radii by \citet{mr88}, we find that
the mass and radius of component~1, which follow from our solution,
also correspond to a normal star of spectral type B9.5-A0\,V. The range of
$M_{\rm bol}=1$\m$01-1$\m18 for it implies $M_{\rm V}=1$\m$19-1$\m36 and
a distance range of $139-151$~pc. Our combined solution thus agrees with
the observed distance of \bn.
In passing, we also note that the effective temperature of component~2,
based on the comparison of model and disentangled spectra, 4655~K, corresponds
to spectral type G8~III according to a calibration by \citet{popper80}.
The mass of component~2 is, however, much lower than the mass of
a normal G8 giant, as usually found for the mass-losing stars in binaries.

\section{\br among other known emission-line semi-detached binaries}
As already mentioned, the number of Be stars with known binary companions
has been increasing steadily. However, many of them are single-line
spectroscopic binaries and the nature of their companions is not known.
There are some visual binaries among Be stars \citep{oud2010}.
It is not clear how strong the mutual interaction between the components
of such systems can be,
although there is a suspicion of strong interaction during periastron
passages in some systems with highly eccentric orbits like $\delta$~Sco
\citep[cf.][and references therein]{mei2013}. Another distinct subgroup
of Be binaries are systems with hot and compact secondaries, the prototype
of such systems being $\varphi$~Per \citep{gies98,bozic95}. Only five such systems
are known \citep[see][and references therein]{kou2012,kou2014}.
Finally, a subgroup of mass-exchanging semi-detached binaries exists in
which the Be components are the mass-gaining components as predicted by
the original binary hypothesis of the Be phenomenon \citep{bebin75,bebin76}.

There has also been recent progress in modelling the process of mass
exchange in binaries \citep[e.g.][]{siess2013,des2013,davis2014} and
obtaining reliable physical properties of mass-exchanging binaries is
therefore essential for testing the predictions of new models. In
Table~\ref{emisbin}, we collected a list of well-studied semi-detached
emission-line binaries known to us and their published basic properties.
We note that the spectral types quoted for mass gainers refer to the stellar
bodies, not to the observed spectral classes affected by circumstellar matter.
A zero secular change of the orbital period means that no significant
change was detected in the available data.

The systems similar to those listed in Table~\ref{emisbin} may not actually
be very rare. We included only systems with reasonably safe evidence that one
of the stars is filling its Roche lobe. The same might be true for the systems
listed in Table~\ref{puta}, for which complete orbital and/or light-curve
solutions and information on a secular orbital period change are still
missing. Their detailed studies are very desirable.
(Here, we have not included into consideration a number of short-periodic
semi-detached binaries, for which transient \ha emission was reported
in the astronomical literature.)

The list in Table~\ref{emisbin} is too short to be statistically significant,
but we use it to point out that there still might be a long way to a detailed
comparison between observations and theory:
\begin{enumerate}
\item One important observable is the equatorial rotational velocity
of the gainer, but it is not often available even in new studies. For instance,
\citet{men2012a} considered two models in their study of V393~Sco:
spin-orbit synchonisation of the gainer and gainer with a break-up velocity.
Available line profiles of the gainer show that its \vsin is somewhere in
between these two cases, but this piece of information was not used.
Our present study of \br is probably one of the few where the synchronicity
parameter $F$ was changed after each iteration to correspond to the observed
\vsin of the gainer.
\item It is not obvious why some of these systems exhibit cyclic (if not
periodic) light and colour changes on a time scale that is an order of magnitude
longer than their respective orbital periods. For three of such variables,
$\beta$~Lyr, AU~Mon, and V393~Sco, the presence of bipolar jets was considered.
Similar jets were also considered for TT Hya, so a study of its long-term
photometric behaviour would be of interest.  Non-orbital light changes
were also observed for UX~Mon, but no clear periodicity was found.
It is remarkable that UX~Mon is a rare system at the early phase of the
mass transfer before the mass ratio reversal.  The differences in the orbital
inclination also do not provide a clue. \br and CX~Dra are the only two
non-eclipsing binaries in the considered sample, and while \br seems to be
secularly constant, CX~Dra exhibits rather strong non-orbital long-term
changes.
\item \citet{des2013} tried to classify semi-detached binaries into
three classes: I. systems before the mass ratio reversal; II. systems
after the reversal but with still high rate of mass transfer; and
III. systems in the quiescent final stages of mass transfer. We warn that
the real situation may be even more complex and note that the secular
period decrease would qualify SX~Cas as a candidate for class~I,
while its observed masses clearly show that this is a system after
mass reversal. Obviously, secular period changes may also be caused by
other effects than the mass transfer between the components, for instance,
by the presence of other bodies in the system. We can only conclude that
the results of our study provide solid evidence that \br is a system
at the quiescent stages of mass transfer after the mass ratio reversal.
It has the lowest mass ratio of all systems of Table~\ref{emisbin}.
\end{enumerate}

\begin{table}
\caption{Other possibly semi-detached emission-line binaries.}\label{puta}
\begin{tabular}{rrcrcccrc}
\hline\hline\noalign{\smallskip}
  Binary & HD/ &$P_{\rm orb}$&Principal\\
         &CoD  &     (d)     &references\\
\noalign{\smallskip}\hline\noalign{\smallskip}
         V617 Aur &  37453           &  $66\fd75$   &1\\
         V395~Aur &  43246           &  $23\fd17$   &2\\
           HZ CMa &  50123           &  $28\fd601$  &3\\
        HIP 33657 &  51956           &  $107\fd4$   &4\\
BD-17$^\circ$2010 &  59771           & $\sim90\fd$  &5\\
           PW Pup & -30$^\circ$5135  &  $158\fd0$   &5\\
           BY Cru & 104901           &  $106\fd4$   &6\\
           HL Lib & 127208           &  $24\fd615$  &7\\
        V1914 Cyg & 207739           &  $140\fd782$ &8\\
           KX~And & 218393           &  $38\fd9$    &9\\
\hline\noalign{\smallskip}
\end{tabular}
\tablefoot{
1...\citet{parsons88},
2...\citet{dempsey90},
3...\citet{sterken94},
4...\citet{burki83,parsons93},
5...\citet{eggen83,bopp91},
6...\citet{daems97},
7...\citet{dempsey90},
8...\citet{griffin90,szabados90},
9...\citet{zarfin14,tolja98}\,.
}
\end{table}

\begin{acknowledgements}
This study uses spectral observations obtained with the HERMES
spectrograph, which is supported by the Fund for Scientific Research of
Flanders (FWO), Belgium, the Research Council of K.U.Leuven, Belgium,
the Fonds National de la Recherche Scientifique (F.R.S.-FNRS), Belgium,
the Royal Observatory of Belgium, the Observatoire de Gen\`eve, Switzerland
and the Th\"uringer Landessternwarte Tautenburg, Germany, spectral observations
with the Elodie spectrograph attached to 1.9-m reflector of Haute Provence
Observatory, CCD spectrograph of the Ond\v{r}ejov 2-m reflector, and
CCD spectra from the spectrograph attached to the 1.22-m reflector
of the Dominion Astrophysical Observatory. It also uses photometric
observations made at the observatory of La Palma (Mercator),
Hvar, and South African Astronomical Observatory (SAAO),
and by the ESA Hipparcos satellite.
We gratefully acknowledge the use of spectrograms of \br from the public
archives of the Elodie spectrograph of the Haute Provence Observatory and
the use of the latest publicly available versions of the programs \fotel and
\korel written by Dr. P.~Hadrava. Our sincere thanks are also due
to Dr.~A.Pr\v{s}a, who provided us with a modified version of
the program \phoebe 1.0 and frequent consultations on its usage.
We thank Drs. M.~Ceniga, A.~Kawka, J.~Kub\'at, P.~Mayer,
P.~Nemeth, M.~Netolick\'y, J.~Polster, and V.~Votruba,
who obtained several Ond\v{r}ejov spectra used in this study and to
Dr.~K.~Uytterhoeven, who participated in photometric observations at La~Palma.
The comments of an anonymous referee helped to shorten the text and
improve the presentation of the main results a great deal.
This research was supported by the grants 205/06/0304,
205/08/H005, and P209/10/0715 of the Czech Science Foundation,
by the grant 678212 of the Grant Agency of the Charles University in Prague,
from the research project AV0Z10030501 of the Academy of Sciences of
the Czech Republic, and from
the Research Program MSM0021620860 {\sl Physical study of objects and
processes in the solar system and in astrophysics} of the Ministry
of Education of the Czech Republic. The research of PK was supported
from the ESA PECS grant 98058. HB acknowledges financial support from
the Croatian Science Foundation under the project 6212 ``Solar and
Stellar Variability". PN thanks the Swiss National Science Foundation for
its support in acquiring photometric data in the Geneva 7-C system.
We acknowledge the use of the electronic database from the CDS, Strasbourg, and
the electronic bibliography maintained by the NASA/ADS system.
\end{acknowledgements}

\bibliographystyle{aa}
\bibliography{citace}

\begin{thebibliography}{85}
\expandafter\ifx\csname natexlab\endcsname\relax\def\natexlab#1{#1}\fi

\bibitem[{{Ak} {et~al.}(2007){Ak}, {Chadima}, {Harmanec}, {Demircan}, {Yang},
  {Koubsk{\'y}}, {{\v S}koda}, {{\v S}lechta}, {Wolf}, {Bo{\v z}i{\'c}}, {Ru{\v
  z}djak}, \& {Sudar}}]{ak2007}
{Ak}, H., {Chadima}, P., {Harmanec}, P., {et~al.} 2007, \aap, 463, 233

\bibitem[{{Andersen} {et~al.}(1988){Andersen}, {Nordstrom}, {Mayor}, \&
  {Polidan}}]{ander88}
{Andersen}, J., {Nordstrom}, B., {Mayor}, M., \& {Polidan}, R.~S. 1988, \aap,
  207, 37

\bibitem[{{Andersen} {et~al.}(1989){Andersen}, {Pavlovski}, \&
  {Piirola}}]{ander89}
{Andersen}, J., {Pavlovski}, K., \& {Piirola}, V. 1989, \aap, 215, 272

\bibitem[{{Atwood-Stone} {et~al.}(2012){Atwood-Stone}, {Miller}, {Richards},
  {Budaj}, \& {Peters}}]{atw2012}
{Atwood-Stone}, C., {Miller}, B.~P., {Richards}, M.~T., {Budaj}, J., \&
  {Peters}, G.~J. 2012, \apj, 760, 134

\bibitem[{{Baranne} {et~al.}(1996){Baranne}, {Queloz}, {Mayor}, {Adrianzyk},
  {Knispel}, {Kohler}, {Lacroix}, {Meunier}, {Rimbaud}, \& {Vin}}]{barrane96}
{Baranne}, A., {Queloz}, D., {Mayor}, M., {et~al.} 1996, \aaps, 119, 373

\bibitem[{{Bisikalo} {et~al.}(2000){Bisikalo}, {Harmanec}, {Boyarchuk},
  {Kuznetsov}, \& {Hadrava}}]{bis2000}
{Bisikalo}, D.~V., {Harmanec}, P., {Boyarchuk}, A.~A., {Kuznetsov}, O.~A., \&
  {Hadrava}, P. 2000, \aap, 353, 1009

\bibitem[{{Bopp} {et~al.}(1991){Bopp}, {Dempsey}, \& {Parsons}}]{bopp91}
{Bopp}, B.~W., {Dempsey}, R.~C., \& {Parsons}, S.~B. 1991, \pasp, 103, 444

\bibitem[{{Bo\v{z}i\'c} {et~al.}(1995){Bo\v{z}i\'c}, {Harmanec}, {Horn},
  {Koubsk\'y}, {Scholz}, {McDavid}, {Hubert}, \& {Hubert}}]{bozic95}
{Bo\v{z}i\'c}, H., {Harmanec}, P., {Horn}, J., {et~al.} 1995, \aap, 304, 235

\bibitem[{{Briot} \& {Royer}(2009)}]{briot2009}
{Briot}, D. \& {Royer}, F. 2009, Be Star Newsletter No., 39, 15

\bibitem[{{Burki} \& {Mayor}(1983)}]{burki83}
{Burki}, G. \& {Mayor}, M. 1983, \aap, 124, 256

\bibitem[{{Claret}(1998)}]{claret98}
{Claret}, A. 1998, \aaps, 131, 395

\bibitem[{{Claret}(2001)}]{claret2001}
{Claret}, A. 2001, \mnras, 327, 989

\bibitem[{{Daems} {et~al.}(1997){Daems}, {Waelkens}, \& {Mayor}}]{daems97}
{Daems}, K., {Waelkens}, C., \& {Mayor}, M. 1997, \aap, 317, 823

\bibitem[{{David} {et~al.}(2014){David}, {Blomme}, {Fr{\'e}mat}, {Damerdji},
  {Delle Luche}, {Gosset}, {Katz}, \& {Viala}}]{david2014}
{David}, M., {Blomme}, R., {Fr{\'e}mat}, Y., {et~al.} 2014, \aap, 562, A97

\bibitem[{{David} \& {Verschueren}(1995)}]{david95}
{David}, M. \& {Verschueren}, W. 1995, \aaps, 111, 183

\bibitem[{{Davis} {et~al.}(2014){Davis}, {Siess}, \& {Deschamps}}]{davis2014}
{Davis}, P.~J., {Siess}, L., \& {Deschamps}, R. 2014, \aap, 570, A25

\bibitem[{{de Laverny} {et~al.}(2012){de Laverny}, {Recio-Blanco}, {Worley}, \&
  {Plez}}]{delaverny2012}
{de Laverny}, P., {Recio-Blanco}, A., {Worley}, C.~C., \& {Plez}, B. 2012,
  \aap, 544, A126

\bibitem[{{Dempsey} {et~al.}(1990){Dempsey}, {Bopp}, {Parsons}, \&
  {Fekel}}]{dempsey90}
{Dempsey}, R.~C., {Bopp}, B.~W., {Parsons}, S.~B., \& {Fekel}, F.~C. 1990,
  \pasp, 102, 312

\bibitem[{{Deschamps} {et~al.}(2013){Deschamps}, {Siess}, {Davis}, \&
  {Jorissen}}]{des2013}
{Deschamps}, R., {Siess}, L., {Davis}, P.~J., \& {Jorissen}, A. 2013, \aap,
  557, A40

\bibitem[{{Desmet} {et~al.}(2010){Desmet}, {Fr{\'e}mat}, {Baudin}, {Harmanec},
  {Lampens}, {Pacheco}, {Briquet}, {Degroote}, {Neiner}, {Mathias}, {Poretti},
  {Rainer}, {Uytterhoeven}, {Amado}, {Valtier}, {Pr{\v s}a}, {Maceroni}, \&
  {Aerts}}]{desmet2010}
{Desmet}, M., {Fr{\'e}mat}, Y., {Baudin}, F., {et~al.} 2010, \mnras, 401, 418

\bibitem[{{Djura{\v s}evi\'c} {et~al.}(2009){Djura{\v s}evi\'c}, {Vince},
  {Khruzina}, \& {Rovithis-Livaniou}}]{djuras2009}
{Djura{\v s}evi\'c}, G., {Vince}, I., {Khruzina}, T.~S., \&
  {Rovithis-Livaniou}, E. 2009, \mnras, 396, 1553

\bibitem[{{Dubath} {et~al.}(2011){Dubath}, {Rimoldini}, {S{\"u}veges},
  {Blomme}, {L{\'o}pez}, {Sarro}, {De Ridder}, {Cuypers}, {Guy}, {Lecoeur},
  {Nienartowicz}, {Jan}, {Beck}, {Mowlavi}, {De Cat}, {Lebzelter}, \&
  {Eyer}}]{dub2011}
{Dubath}, P., {Rimoldini}, L., {S{\"u}veges}, M., {et~al.} 2011, \mnras, 414,
  2602

\bibitem[{{Eaton}(2008)}]{eaton2008}
{Eaton}, J.~A. 2008, \apj, 681, 562

\bibitem[{{Eggen}(1983)}]{eggen83}
{Eggen}, O.~J. 1983, \aj, 88, 1676

\bibitem[{{Gies} {et~al.}(1998){Gies}, {Bagnuolo}, {Ferrara}, {Kaye},
  {Thaller}, {Penny}, \& {Peters}}]{gies98}
{Gies}, D.~R., {Bagnuolo}, Jr., W.~G., {Ferrara}, E.~C., {et~al.} 1998, \apj,
  493, 440

\bibitem[{{Griffin} {et~al.}(1990){Griffin}, {Parsons}, {Dempsey}, \&
  {Bopp}}]{griffin90}
{Griffin}, R.~F., {Parsons}, S.~B., {Dempsey}, R., \& {Bopp}, B.~W. 1990,
  \pasp, 102, 535

\bibitem[{{Hadrava}(1990)}]{fotel1}
{Hadrava}, P. 1990, Contr. Astron. Obs. Skalnat\'e Pleso, 20, 23

\bibitem[{{Hadrava}(1995)}]{korel1}
{Hadrava}, P. 1995, \aaps, 114, 393

\bibitem[{{Hadrava}(2004{\natexlab{a}})}]{fotel2}
{Hadrava}, P. 2004{\natexlab{a}}, Publ. Astron. Inst. Acad. Sci. Czech Rep.,
  92, 1

\bibitem[{{Hadrava}(2004{\natexlab{b}})}]{korel3}
{Hadrava}, P. 2004{\natexlab{b}}, Publ. Astron. Inst. Acad. Sci. Czech Rep.,
  92, 15

\bibitem[{{Harmanec}(1988)}]{mr88}
{Harmanec}, P. 1988, Bulletin of the Astronomical Institutes of Czechoslovakia,
  39, 329

\bibitem[{{Harmanec}(1998)}]{hpvb}
{Harmanec}, P. 1998, \aap, 335, 173

\bibitem[{{Harmanec}(2002)}]{hec2002}
{Harmanec}, P. 2002, AN, 323, 87

\bibitem[{{Harmanec}(2003)}]{bbin2003}
{Harmanec}, P. 2003, in Publ. Canakkale Onsekiz Mart Univ. Vol. 3: New
  Directions for Close Binary Studies: The Royal Road to the Stars, 221--233

\bibitem[{{Harmanec} \& {Horn}(1998)}]{hechor98}
{Harmanec}, P. \& {Horn}, J. 1998, Journal of Astronomical Data, 4, 5

\bibitem[{{Harmanec} {et~al.}(1994){Harmanec}, {Horn}, \& {Juza}}]{hhj94}
{Harmanec}, P., {Horn}, J., \& {Juza}, K. 1994, \aaps, 104, 121

\bibitem[{{Harmanec} \& {K\v{r}{\'\i}\v{z}}(1976)}]{bebin76}
{Harmanec}, P. \& {K\v{r}{\'\i}\v{z}}, S. 1976, in IAU Symposium, Vol.~70, Be
  and Shell Stars, ed. A.~{Slettebak}, 385

\bibitem[{{Harmanec} {et~al.}(1996){Harmanec}, {Morand}, {Bonneau}, {Jiang},
  {Yang}, {Guinan}, {Hall}, {Mourard}, {Hadrava}, {Bozic}, {Sterken},
  {Tallon-Bosc}, {Walker}, {McCook}, {Vakili}, {Stee}, \& {Le Contel}}]{hec96}
{Harmanec}, P., {Morand}, F., {Bonneau}, D., {et~al.} 1996, \aap, 312, 879

\bibitem[{{Harmanec} \& {Scholz}(1993)}]{hs93}
{Harmanec}, P. \& {Scholz}, G. 1993, \aap, 279, 131

\bibitem[{{Hill} {et~al.}(1997){Hill}, {Harmanec}, {Pavlovski}, {Bo\v{z}i\'c},
  {Hadrava}, {Koubsk\'y}, \& {\v{Z}i\v{z}\v{n}ovsk\'y}}]{zarf17}
{Hill}, G., {Harmanec}, P., {Pavlovski}, K., {et~al.} 1997, \aap, 324, 965

\bibitem[{{Horn} {et~al.}(1996){Horn}, {Kub\'at}, {Harmanec}, {Koubsk\'y},
  {Hadrava}, {\v{S}imon}, {\v{S}tefl}, \& {\v{S}koda}}]{sef0}
{Horn}, J., {Kub\'at}, J., {Harmanec}, P., {et~al.} 1996, \aap, 309, 521

\bibitem[{{Kazarovets} {et~al.}(1999){Kazarovets}, {Samus}, {Durlevich},
  {Frolov}, {Antipin}, {Kireeva}, \& {Pastukhova}}]{kaz99}
{Kazarovets}, E.~V., {Samus}, N.~N., {Durlevich}, O.~V., {et~al.} 1999,
  Information Bulletin on Variable Stars, 4659, 1

\bibitem[{{Koubsk\'y} {et~al.}(1998){Koubsk\'y}, {Harmanec}, {Bo\v{z}i\'c},
  {Percy}, {\v{Z}i\v{z}\v{n}ovsk\'y}, {Huang}, {Richards}, {Hadrava}, \&
  {\v{S}imon}}]{kou98}
{Koubsk\'y}, P., {Harmanec}, P., {Bo\v{z}i\'c}, H., {et~al.} 1998, Hvar
  Observatory Bulletin, 22, 17

\bibitem[{{Koubsk{\'y}} {et~al.}(2014){Koubsk{\'y}}, {Kotkov{\'a}}, {Kraus},
  {Yang}, {{\v S}lechta}, {Harmanec}, {Wolf}, {Votruba}, {Kub\'at},
  {Kub\'atov\'a}, {Niemczura}, \& {{\v S}koda}}]{kou2014}
{Koubsk{\'y}}, P., {Kotkov{\'a}}, L., {Kraus}, M., {et~al.} 2014, \aap, 567,
  A57

\bibitem[{{Koubsk{\'y}} {et~al.}(2012){Koubsk{\'y}}, {Kotkov{\'a}}, {Votruba},
  {{\v S}lechta}, \& {Dvo{\v r}{\'a}kov{\'a}}}]{kou2012}
{Koubsk{\'y}}, P., {Kotkov{\'a}}, L., {Votruba}, V., {{\v S}lechta}, M., \&
  {Dvo{\v r}{\'a}kov{\'a}}, {\v S}. 2012, \aap, 545, A121

\bibitem[{{K\v{r}\'\i\v{z}} \& {Harmanec}(1975)}]{bebin75}
{K\v{r}\'\i\v{z}}, S. \& {Harmanec}, P. 1975, BAICz, 26, 65

\bibitem[{{Linnell} {et~al.}(2006){Linnell}, {Harmanec}, {Koubsk{\'y}}, {Bo{\v
  z}i{\'c}}, {Yang}, {Ru{\v z}djak}, {Sudar}, {Libich}, {Eenens}, {Krpata},
  {Wolf}, {{\v S}koda}, \& {{\v S}lechta}}]{zarf24}
{Linnell}, A.~P., {Harmanec}, P., {Koubsk{\'y}}, P., {et~al.} 2006, \aap, 455,
  1037

\bibitem[{{Lucy} \& {Sweeney}(1971)}]{lucy71}
{Lucy}, L.~B. \& {Sweeney}, M.~A. 1971, \aj, 76, 544

\bibitem[{{Meilland} {et~al.}(2013){Meilland}, {Stee}, {Spang}, {Malbet},
  {Massi}, \& {Schertl}}]{mei2013}
{Meilland}, A., {Stee}, P., {Spang}, A., {et~al.} 2013, \aap, 550, L5

\bibitem[{{Mennickent} {et~al.}(2012{\natexlab{a}}){Mennickent}, {Djura{\v
  s}evi{\'c}}, {Ko{\l}aczkowski}, \& {Michalska}}]{men2012a}
{Mennickent}, R.~E., {Djura{\v s}evi{\'c}}, G., {Ko{\l}aczkowski}, Z., \&
  {Michalska}, G. 2012{\natexlab{a}}, \mnras, 421, 862

\bibitem[{{Mennickent} {et~al.}(2012{\natexlab{b}}){Mennickent},
  {Ko{\l}aczkowski}, {Djura{\v s}evi\'c}, {Niemczura}, {Diaz}, {Cur{\'e}},
  {Araya}, \& {Peters}}]{men2012b}
{Mennickent}, R.~E., {Ko{\l}aczkowski}, Z., {Djura{\v s}evi\'c}, G., {et~al.}
  2012{\natexlab{b}}, \mnras, 427, 607

\bibitem[{{Miller} {et~al.}(2007){Miller}, {Budaj}, {Richards}, {Koubsk{\'y}},
  \& {Peters}}]{miller2007}
{Miller}, B., {Budaj}, J., {Richards}, M., {Koubsk{\'y}}, P., \& {Peters},
  G.~J. 2007, \apj, 656, 1075

\bibitem[{{Moore}(1945)}]{moore45}
{Moore}, C.~E. 1945, Contributions from the Princeton University Observatory,
  20, D23

\bibitem[{{Nasseri} {et~al.}(2014){Nasseri}, {Chini}, {Harmanec}, {Mayer},
  {Nemravov{\'a}}, {Dembsky}, {Lehmann}, {Sana}, \& {Le Bouquin}}]{anita2014}
{Nasseri}, A., {Chini}, R., {Harmanec}, P., {et~al.} 2014, \aap, 568, A94

\bibitem[{{Olson} {et~al.}(2009){Olson}, {Henry}, \& {Etzel}}]{olson2009}
{Olson}, E.~C., {Henry}, G.~W., \& {Etzel}, P.~B. 2009, \aj, 138, 1435

\bibitem[{{Oudmaijer} \& {Parr}(2010)}]{oud2010}
{Oudmaijer}, R.~D. \& {Parr}, A.~M. 2010, \mnras, 405, 2439

\bibitem[{{Palacios} {et~al.}(2010){Palacios}, {Gebran}, {Josselin}, {Martins},
  {Plez}, {Belmas}, \& {Lèbre}}]{palacios2010}
{Palacios}, A., {Gebran}, M., {Josselin}, E., {et~al.} 2010, \aap, 516, A13

\bibitem[{{Parsons} \& {Bopp}(1993)}]{parsons93}
{Parsons}, S.~B. \& {Bopp}, B.~W. 1993, in Bulletin of the American
  Astronomical Society, Vol.~25, American Astronomical Society Meeting
  Abstracts, 1376

\bibitem[{{Parsons} {et~al.}(1988){Parsons}, {Dempsey}, \& {Bopp}}]{parsons88}
{Parsons}, S.~B., {Dempsey}, R.~C., \& {Bopp}, B.~W. 1988, in ESA Special
  Publication, Vol. 281, ESA Special Publication, 225--227

\bibitem[{{Paunzen} {et~al.}(2001){Paunzen}, {Duffee}, {Heiter}, {Kuschnig}, \&
  {Weiss}}]{paun2001}
{Paunzen}, E., {Duffee}, B., {Heiter}, U., {Kuschnig}, R., \& {Weiss}, W.~W.
  2001, \aap, 373, 625

\bibitem[{{Perryman} \& {ESA}(1997)}]{esa97}
{Perryman}, M.~A.~C. \& {ESA}. 1997, {The HIPPARCOS and TYCHO catalogues}
  (Astrometric and photometric star catalogues derived from the ESA Hipparcos
  Space Astrometry Mission, Publisher: Noordwijk, Netherlands: ESA Publications
  Division, 1997, Series: ESA SP Series 1200)

\bibitem[{{Peters} {et~al.}(2013){Peters}, {Pewett}, {Gies}, {Touhami}, \&
  {Grundstrom}}]{peters2013}
{Peters}, G.~J., {Pewett}, T.~D., {Gies}, D.~R., {Touhami}, Y.~N., \&
  {Grundstrom}, E.~D. 2013, \apj, 765, 2

\bibitem[{{Pojmanski}(2002)}]{pojm2002}
{Pojmanski}, G. 2002, Acta Astronomica, 52, 397

\bibitem[{{Popper}(1980)}]{popper80}
{Popper}, D.~M. 1980, \araa, 18, 115

\bibitem[{Press {et~al.}(1993)Press, Teukolsky, Vetterling, \&
  Flannery}]{press93}
Press, W.~H., Teukolsky, S.~A., Vetterling, W.~T., \& Flannery, B.~P. 1993,
  Numerical Recipes in FORTRAN; The Art of Scientific Computing, 2nd edn. (New
  York, NY, USA: Cambridge University Press)

\bibitem[{{Pr{\v s}a} \& {Zwitter}(2005)}]{prsa2005}
{Pr{\v s}a}, A. \& {Zwitter}, T. 2005, \apj, 628, 426

\bibitem[{{Pr{\v s}a} \& {Zwitter}(2006)}]{prsa2006}
{Pr{\v s}a}, A. \& {Zwitter}, T. 2006, \apss, 36

\bibitem[{{Raskin} {et~al.}(2004){Raskin}, {Burki}, {Burnet}, {Davignon},
  {Dubosson}, {Ischi}, {George}, {Grenon}, {Maire}, {Van Winckel}, {Waelkens},
  \& {Weber}}]{raskin}
{Raskin}, G., {Burki}, G., {Burnet}, M., {et~al.} 2004, in SPIE Conf. Series,
  Vol. 5492, Ground-based Instrumentation for Astronomy, ed. A.~F.~M.
  {Moorwood} \& M.~{Iye}, 830--840

\bibitem[{{Raskin} {et~al.}(2011){Raskin}, {van Winckel}, {Hensberge},
  {Jorissen}, {Lehmann}, {Waelkens}, {Avila}, {de Cuyper}, {Degroote},
  {Dubosson}, {Dumortier}, {Fr{\'e}mat}, {Laux}, {Michaud}, {Morren}, {Perez
  Padilla}, {Pessemier}, {Prins}, {Smolders}, {van Eck}, \&
  {Winkler}}]{raskin2011}
{Raskin}, G., {van Winckel}, H., {Hensberge}, H., {et~al.} 2011, \aap, 526, A69

\bibitem[{{Renson} \& {Manfroid}(2009)}]{rens2009}
{Renson}, P. \& {Manfroid}, J. 2009, \aap, 498, 961

\bibitem[{{Richards} {et~al.}(2000){Richards}, {Koubsk{\'y}}, {{\v S}imon},
  {Peters}, {Hirata}, {{\v S}koda}, \& {Masuda}}]{rich2000}
{Richards}, M.~T., {Koubsk{\'y}}, P., {{\v S}imon}, V., {et~al.} 2000, \apj,
  531, 1003

\bibitem[{{Royer} {et~al.}(2007){Royer}, {Briot}, {North}, {Burki}, \&
  {Carrier}}]{roy2007}
{Royer}, F., {Briot}, D., {North}, P., {Burki}, G., \& {Carrier}, F. 2007, in
  IAU Symposium, Vol. 240, IAU Symposium, ed. {W.~I.~Hartkopf, E.~F.~Guinan, \&
  P.~Harmanec}, 211

\bibitem[{{Ru{\v z}djak} {et~al.}(2009){Ru{\v z}djak}, {Bo{\v z}i{\'c}},
  {Harmanec}, {Fi{\v r}t}, {Chadima}, {Bjorkman}, {Gies}, {Kaye},
  {Koubsk{\'y}}, {McDavid}, {Richardson}, {Sudar}, {{\v S}lechta}, {Wolf}, \&
  {Yang}}]{zarf26}
{Ru{\v z}djak}, D., {Bo{\v z}i{\'c}}, H., {Harmanec}, P., {et~al.} 2009, \aap,
  506, 1319

\bibitem[{{Siess} {et~al.}(2013){Siess}, {Izzard}, {Davis}, \&
  {Deschamps}}]{siess2013}
{Siess}, L., {Izzard}, R.~G., {Davis}, P.~J., \& {Deschamps}, R. 2013, \aap,
  550, A100

\bibitem[{{\v{S}koda}(1996)}]{spefo}
{\v{S}koda}, P. 1996, in ASP Conf. Ser. 101: Astronomical Data Analysis
  Software and Systems V, 187--189

\bibitem[{{\v{S}tefl} {et~al.}(1990){\v{S}tefl}, {Harmanec}, {Horn}, {Koubsky},
  {K\v{r}\'\i\v{z}}, {Hadrava}, {Bo\v{z}i\'c}, \& {Pavlovski}}]{zarfin14}
{\v{S}tefl}, S., {Harmanec}, P., {Horn}, J., {et~al.} 1990, BAiCz, 41, 29

\bibitem[{{Sterken} {et~al.}(1994){Sterken}, {Vogt}, \&
  {Mennickent}}]{sterken94}
{Sterken}, C., {Vogt}, N., \& {Mennickent}, R. 1994, \aap, 291, 473

\bibitem[{{Sterne}(1941)}]{sterne41}
{Sterne}, T.~E. 1941, Proc. Nat. Acad. Sci., 27, 168

\bibitem[{{Stetson}(1991)}]{stet91}
{Stetson}, P.~B. 1991, \aj, 102, 589

\bibitem[{{Sudar} {et~al.}(2011){Sudar}, {Harmanec}, {Lehmann}, {Yang}, {Bo{\v
  z}i{\'c}}, \& {Ru{\v z}djak}}]{sudar2011}
{Sudar}, D., {Harmanec}, P., {Lehmann}, H., {et~al.} 2011, \aap, 528, A146

\bibitem[{{Szabados}(1990)}]{szabados90}
{Szabados}, L. 1990, \aap, 232, 381

\bibitem[{{Tarasov} {et~al.}(1998){Tarasov}, {Berdyugina}, \&
  {Berdyugin}}]{tolja98}
{Tarasov}, A.~E., {Berdyugina}, S.~V., \& {Berdyugin}, A.~V. 1998, Astronomy
  Letters, 24, 316

\bibitem[{{van Leeuwen}(2007{\natexlab{a}})}]{leeuw2007b}
{van Leeuwen}, F. 2007{\natexlab{a}}, in Astrophysics and Space Science
  Library, Vol. 350, Astrophysics and Space Science Library, ed. {F.~van
  Leeuwen}

\bibitem[{{van Leeuwen}(2007{\natexlab{b}})}]{leeuw2007a}
{van Leeuwen}, F. 2007{\natexlab{b}}, \aap, 474, 653

\bibitem[{{Zucker} \& {Mazeh}(1994)}]{todcor1}
{Zucker}, S. \& {Mazeh}, T. 1994, \apj, 420, 806

\end{thebibliography}

\Online
\begin{appendix}
\section{Details of the spectral data reduction and measurements}\label{apa}
The initial reduction of all Ond\v{r}ejov and DAO spectra (bias subtraction,
flat-fielding, creation of 1-D spectra, and wavelength calibration) was carried
out in {\tt IRAF}. For Elodie and Hermes spectra, dedicated reduction pipelines
were used, combined with some IDL routines in the case of Elodie.
See \citet{barrane96} and \citet{raskin2011} for detailed descriptions of
the Elodie and Hermes spectrographs.
Rectification, removal of residual cosmics and flaws and
RV measurements of all spectra were carried out with the program \spefo
\citep{sef0,spefo}, namely the latest version 2.63 developed by Mr.~J.~Krpata.
\spefo displays direct and flipped traces of the line
profiles superimposed on the computer screen that the user can slide
to achieve a precise overlapping of the parts of the profile of whose RV
one wants to measure.
Using a selection of stronger unblended lines of the cool component~2
(see Table~\ref{klist}) covering the red spectral region (available for
all spectra), we measured RVs of all of them to obtain a mean RV for
all spectra. We also measured a selection of good telluric lines and used
them for an additional fine correction of the RV zero point of each spectrogram
\citep{sef0}. Moreover, we measured the RVs of the steep wings of the
\ha emission line and of the two absorption cores of \hae. We point out
that although some broad and shallow lines of component~1 are seen
in the spectra, their direct RV measurement is impossible because of
numerous blends with the lines of component~2.

\begin{table}
\caption[]{List of spectral lines of component~2 and their air
wavelengths used for the RV measurements in \spefoe. The numbering of
the multiplets of individual ions corresponds to the one introduced by
\citet{moore45}.}
\label{klist}
\begin{center}
\begin{tabular}{clclcccl}
\hline\hline\noalign{\smallskip}
Wavelength   & Element & Wavelength   & Element \\
(\AA)       &         &  (\AA)      \\
\noalign{\smallskip}\hline\noalign{\smallskip}
 6322.694& \ion{Fe}{i} 207  & 6439.083& \ion{Ca}{i} 18    \\
 6327.604& \ion{Ni}{i} 49   & 6643.638& \ion{Ni}{i} 43    \\
 6355.035& \ion{Fe}{i} 342  & 6677.997& \ion{Fe}{i} 268   \\
 6358.687& \ion{Fe}{i} 13   & 6705.105& \ion{Fe}{i} 1197  \\
 6393.612& \ion{Fe}{i} 168  & 6717.687& \ion{Ca}{i} 32    \\
 6411.658& \ion{Fe}{i} 816  & 6726.673& \ion{Fe}{i} 1197  \\
 6430.856& \ion{Fe}{i} 62   &         &            \\
\noalign{\smallskip}\hline\noalign{\smallskip}
\end{tabular}
\end{center}
\end{table}

\begin{table*}
\caption[]{Individual \spefo RVs of component~2. Letters E, O, D, and H in
column ``Spg." denote the Elodie, Ond\v{r}ejov, DAO, and Hermes
spectrographs.}\label{rvsp}
\begin{flushleft}
\begin{tabular}{crccrccrc}
\hline\hline\noalign{\smallskip}
HJD       &$RV$ & Spg.&HJD       &$RV$ & Spg.&HJD    &$RV$ & Spg.\\
-2400000  &(\ks)&     &-2400000  &(\ks)&     &-2400000  &(\ks)   \\
\noalign{\smallskip}\hline\noalign{\smallskip}
50071.5415&  17.71& E& 54115.5309&   2.80& O&55592.4540& -98.74& O\\
50072.5331& -27.10& E& 54116.4556& -38.40& O&55593.3881& -82.17& O\\
50527.3026&-101.63& E& 54172.3849& -78.18& O&55601.5399& -10.33& O\\
50528.2947& -98.11& E& 54173.4495& -38.80& O&55602.5009& -52.10& O\\
50529.2966& -76.42& E& 54174.4208&   2.48& O&55613.5249&  29.96& O\\
50530.3328& -37.71& E& 54176.4755&  67.45& O&55614.4452&  -7.48& O\\
50842.5007&  46.40& E& 54185.3680& -74.68& O&55621.4578&   6.53& O\\
50843.6757&  72.95& E& 54185.3914& -74.08& O&55623.4517&  67.26& O\\
51565.4951&  30.44& E& 54185.4381& -74.44& O&55628.4816& -59.02& O\\
51567.4662&  76.03& E& 54186.3577& -39.80& O&55644.4341& -96.59& O\\
52215.6544&  34.01& E& 54193.4186& -13.47& O&55645.4161& -70.53& O\\
52216.6342&  -5.82& E& 54197.3561& -98.74& O&55649.4090&  70.69& O\\
52217.6902& -51.60& E& 54205.3780&  25.44& O&54733.0084&  74.78& D\\
52220.6997& -97.73& E& 54206.4006& -15.76& O&54939.6777&  74.31& D\\
53007.5331&-100.19& E& 54211.3746& -69.48& O&54939.7269&  74.40& D\\
53008.5693&-101.34& E& 54222.3531&-104.84& O&54942.6759& -11.64& D\\
53009.5262& -82.21& E& 54384.5600&  71.98& O&55469.0305&  74.34& D\\
53010.5413& -47.46& E& 54388.6316& -71.52& O&55471.0483&  41.72& D\\
53011.6043&  -3.39& E& 54508.4493& -74.42& O&55570.0131&  15.43& D\\
53014.4313&  73.96& E& 54535.4177& -30.08& O&55501.0092&-101.26& D\\
53413.5151&  50.38& E& 54569.3620& -67.56& O&55601.9629& -29.49& D\\
53414.4822&  71.49& E& 54840.5091& -61.06& O&55602.9492& -68.74& D\\
53416.3923&  57.77& E& 54932.3291& -98.41& O&55858.0824&  58.42& D\\
53418.4050& -19.47& E& 54934.3986& -83.54& O&55953.9157&-101.92& D\\
53419.4041& -61.42& E& 54937.3221&  27.71& O&55993.7744& -84.07& D\\
53420.3829& -90.60& E& 55274.4052&  64.25& O&55994.7681& -50.46& D\\
53421.3983&-102.84& E& 55278.2936&   0.91& O&56031.7031& -99.73& D\\
53422.4619& -93.35& E& 55280.3356& -81.24& O&56296.0462&  74.91& D\\
53424.4110& -28.41& E& 55281.4343&-102.20& O&56381.7600& -71.06& D\\
53425.4095&  13.94& E& 55293.3180& -82.46& O&56382.7638& -32.61& D\\
53426.4032&  50.64& E& 55294.3667&-100.89& O&56383.7508&   9.30& D\\
53428.4087&  72.95& E& 55295.3084&-101.04& O&55589.5585& -51.62& H\\
53429.3832&  55.55& E& 55296.3068& -79.51& O&55592.5374& -98.92& H\\
53430.3826&  21.99& E& 55315.3200&  61.49& O&55595.4975&   0.26& H\\
54000.6311& -56.47& O& 55480.6300&  54.66& O&55904.7369& -34.63& H\\
54017.6633& -68.43& O& 55491.5525& -24.09& O&55906.6854&  43.68& H\\
54018.6197& -31.83& O& 55578.4610&-100.52& O&55908.6513&  74.17& H\\
54019.5490&   7.60& O& 55579.5032&-100.38& O&55910.6683&  28.73& H\\
54025.5356& -17.63& O& 55590.4031& -81.03& O&55912.7053& -58.62& H\\
\noalign{\smallskip}\hline\noalign{\smallskip}
\end{tabular}
\end{flushleft}
\end{table*}

\begin{table*}
\caption[]{Individual {\tt asTODCOR} RVs of both components.
The weights are proportional to the square of the S/N
of individual spectra.}\label{rvtd}
\begin{center}
\begin{tabular}{crrccrrcccrr}
\hline\hline\noalign{\smallskip}
HJD     &$RV_1$ & $RV_2$&weight&HJD     &$RV_1$ & $RV_2$&weight\\
-2400000&(\ks)  & (\ks) &      &-2400000& (\ks) & (\ks) &     \\
\noalign{\smallskip}\hline\noalign{\smallskip}
50071.5415&-1.160& 31.356&0.538&53416.3923&-4.494& 71.609&0.424\\
50072.5331& 1.731&-12.346&0.842&53418.4050& 0.610& -4.722&0.968\\
50527.3026& 4.675&-87.466&0.840&53419.4041& 3.331&-46.735&0.765\\
50528.2947& 5.177&-83.880&0.655&53420.3829& 4.181&-76.069&0.988\\
50529.2966& 4.325&-60.928&0.706&53421.3983& 4.300&-88.491&0.915\\
50530.3328& 1.951&-22.894&1.333&53422.4619& 4.515&-78.451&0.426\\
50842.5007&-3.226& 60.359&1.302&53424.4110& 0.773&-13.666&0.688\\
51565.4951&-2.191& 43.976&0.503&53425.4095&-2.012& 28.347&0.680\\
51567.4662&-4.117& 88.335&0.440&53426.4032&-2.985& 63.936&0.511\\
52215.6544&-3.967& 48.493&0.858&53428.4087&-5.194& 86.918&0.872\\
52216.6342&-1.243&  8.344&1.335&53429.3832&-4.271& 69.230&0.318\\
52217.6902& 1.581&-37.213&0.781&53430.3826&-2.161& 36.197&0.850\\
52220.6997& 5.196&-83.691&0.497&55589.5585& 3.450&-37.172&1.578\\
53007.5331& 4.374&-85.632&1.310&55592.5374& 6.050&-84.350&0.840\\
53008.5693& 4.292&-86.321&0.761&55595.4975&-0.645& 14.860&1.345\\
53009.5262& 4.938&-67.760&1.824&55904.7369& 1.693&-19.898&1.590\\
53010.5413& 2.655&-32.753&1.185&55906.6854&-3.181& 57.918&1.406\\
53011.6043&-1.166& 11.645&1.440&55908.6513&-4.222& 88.134&1.124\\
53014.4313&-5.432& 88.287&1.501&55910.6683&-3.695& 42.479&1.566\\
53413.5151&-3.674& 65.533&0.359&55912.7053& 4.324&-43.761&1.270\\
53414.4822&-5.642& 84.895&2.866&                               \\
\noalign{\smallskip}\hline\noalign{\smallskip}
\end{tabular}
\end{center}
\end{table*}

\section{Details on the photometric data reductions}\label{apb}
Since we used photometry from various sources and photometric systems,
both all-sky and differential, relative to several different
comparison stars, we attempted to arrive at some homogenisation and
standardisation. Special effort was made to derive
improved all-sky values for the comparison stars used,
employing carefully standardised \ubv\ observations secured
at Hvar over several decades of systematic observations.
The adopted values are collected in Table~\ref{comp},
together with the number of all-sky observations and the rms errors of
one observation.
They were added to the respective magnitude differences to obtain directly
comparable standard \ubv\ magnitudes for all stations. To illustrate the
accuracy, with which various data sets were transformed to the standard
system, we give (in Table~\ref{checks}) mean differential \ubv\ values for
the check stars used. They were derived relatively
to the Hvar values for the comparison stars HD 58187 and HD 61341.

\begin{table*}
\caption[]{ Accurate Hvar and SAAO all-sky mean \ubv\ values for all comparison stars used. These were
added to the magnitude differences var.$-$comp. and check$-$comp.
for data from all stations.}\label{comp}
\begin{center}
\begin{tabular}{crcrcccrr}
\hline\hline\noalign{\smallskip}
Station&Star& HD & No. of& $V$  &  $B$ &  $U$ &$(B-V)$& $(U-B)$ \\
       &    &    &   obs.&(mag.)&(mag.)&(mag.)&(mag.) &  (mag.) \\
\noalign{\smallskip}\hline\noalign{\smallskip}
01&     1 CMi&58187& 466& 5.393\p0.010& 5.493\p0.011&5.622\p0.012& 0.100& 0.129\\
01&   HR 2858&59059& 364& 6.254\p0.010& 6.206\p0.011&6.103\p0.013&-0.048&-0.103\\
01&SAO 115753&61341&  39& 7.991\p0.010& 8.182\p0.009&8.281\p0.024& 0.191& 0.099\\
\noalign{\smallskip}\hline\noalign{\smallskip}
11&SAO 115753&61341&   3& 7.986\p0.010& 8.178\p0.013&8.288\p0.015& 0.192& 0.110\\
11&SAO 115750& --  &  24& 9.743\p0.011& 9.883\p0.013&10.012\p0.018& 0.139& 0.129\\
\noalign{\smallskip}\hline\noalign{\smallskip}
\end{tabular}
\end{center}
\end{table*}

\begin{table*}
\caption[]{ Mean \ubv\ values for the check stars used at individual
observing stations derived differentially relative to their respective
comparison stars. They illustrate how closely
the data could be transformed to a comparable standard system.}\label{checks}
\begin{center}
\begin{tabular}{crcrcccrr}
\hline\hline\noalign{\smallskip}
Station&Star& HD & No. of& $V$  &  $B$ &  $U$ &$(B-V)$& $(U-B)$ \\
       &    &    &   obs.&(mag.)&(mag.)&(mag.)&(mag.) &  (mag.) \\
\noalign{\smallskip}\hline\noalign{\smallskip}
01&   HR 2858&59059& 542& 6.254\p0.010& 6.206\p0.009&6.102\p0.013&-0.048&-0.104\\
01&SAO 115753&61341&  53& 7.987\p0.013& 8.182\p0.011&8.278\p0.023& 0.195& 0.096\\
\noalign{\smallskip}\hline\noalign{\smallskip}
11&SAO 115750& --  &  28& 9.747\p0.009& 9.886\p0.011&10.004\p0.013& 0.140& 0.118\\
\noalign{\smallskip}\hline\noalign{\smallskip}
\end{tabular}
\end{center}
\end{table*}

Below, we provide some details of the individual data sets and their reductions.
\begin{itemize}
\item {\sl Station 01 -- Hvar:} \ \
 These differential observations have been secured by HB and PZ relative to
HD~58187 (the check star HD~59059 being observed as frequently as the variable)
and carefully transformed to the standard $UBV$ system via non-linear
transformation formul\ae\ using the {\tt HEC22} reduction program -- see
\citet{hhj94} and \citet{hechor98} for the observational strategy and
data reduction. \footnote{The whole program suite with a detailed manual,
examples of data, auxiliary data files, and results is available at
{\sl http://astro.troja.mff.cuni.cz/ftp/hec/PHOT}\,.}
All observations were reduced with the latest
{\tt HEC22 rel.17} program, which allows the time variation of
linear extinction coefficients to be modelled in the course of observing nights.
\item {\sl Station 11 -- South African Astronomical Observatory (SAAO):}
\ \  These differential \ubv\ observations were obtained by PZ with the 0.50-m reflector
relative to HD~61341 (SAO~115750 being used as the check star) and also
transformed to the standard Johnson system with the help of {\tt HEC22}.
\item {\sl Station 37 -- La Palma} \ \ These all-sky seven-colour (7-C)
observations were secured in the Geneva photometric system using the two-channel aperture
photometer P7-2000 \citep{raskin} mounted on the 1.20-m Belgian Mercator
reflector at the Observatorio de los Muchachos at La~Palma.
\item {\sl Station 61 -- Hipparcos:} \ \ These all-sky observations were
reduced to the standard $V$ magnitude via the transformation formul\ae\
derived by \citet{hpvb} to verify that no secular light changes in
the system were observed. However, for the light-curve solution
in \phoebee, we consider the Hipparcos transmission curve for the $H_p$
magnitude.
\item {\sl Station 93 -- ASAS3 V photometry:} \ \ We extracted these
all-sky observations from the ASAS3 public archive \citep{pojm2002},
using the data for diaphragm~1, having on average the lowest rms errors.
We omitted all observations of grade D and observations having rms errors
larger than 0\m04. We also omitted a strongly deviating observation at
HJD~2452662.6863.
\end{itemize}

\begin{table*}
\caption[]{Individual Geneva 7-C observations.}\label{photg}
\begin{flushleft}
\begin{tabular}{ccccccccc}
\hline\hline\noalign{\smallskip}
 HJD       &  $U$ & $B$ &  $V$ &$B1$ &$B2$ & $V1$& $G$ \\
-2400000   &(mag)&(mag)&(mag)&(mag)&(mag)&(mag)&(mag)\\
\noalign{\smallskip}\hline\noalign{\smallskip}
52307.44605& 7.881&6.495& 7.173&7.417&7.940&7.894&8.275\\
52307.55703& 7.876&6.486& 7.159&7.409&7.931&7.876&8.262\\
52308.45517& 7.900&6.497& 7.161&7.426&7.938&7.876&8.265\\
52308.55777& 7.881&6.472& 7.130&7.404&7.904&7.846&8.233\\
52309.43489& 7.892&6.480& 7.126&7.414&7.915&7.847&8.218\\
52309.56210& 7.886&6.474& 7.112&7.403&7.905&7.832&8.210\\
52312.45134& 7.894&6.492& 7.154&7.418&7.935&7.878&8.261\\
52313.53300& 7.907&6.503& 7.171&7.426&7.945&7.890&8.275\\
52317.53358& 7.867&6.455& 7.092&7.387&7.886&7.811&8.185\\
52318.52657& 7.887&6.484& 7.135&7.412&7.918&7.859&8.236\\
52339.42680& 7.900&6.502& 7.177&7.426&7.939&7.893&8.275\\
52350.43584& 7.881&6.475& 7.130&7.400&7.913&7.851&8.221\\
52351.42891& 7.901&6.500& 7.163&7.425&7.934&7.888&8.266\\
52352.42504& 7.899&6.503& 7.178&7.431&7.947&7.898&8.289\\
52353.41762& 7.893&6.497& 7.164&7.423&7.938&7.885&8.268\\
52354.41404& 7.888&6.481& 7.133&7.418&7.918&7.856&8.235\\
52355.41486& 7.885&6.474& 7.108&7.409&7.905&7.832&8.207\\
52362.37901& 7.881&6.470& 7.107&7.400&7.905&7.828&8.201\\
52367.39349& 7.861&6.458& 7.119&7.386&7.901&7.841&8.219\\
52368.39548& 7.880&6.461& 7.095&7.396&7.895&7.818&8.191\\
52369.38697& 7.873&6.461& 7.095&7.392&7.891&7.817&8.193\\
52370.38003& 7.877&6.474& 7.127&7.399&7.908&7.848&8.225\\
52371.39401& 7.897&6.500& 7.165&7.421&7.939&7.885&8.269\\
52384.38727& 7.890&6.486& 7.152&7.411&7.925&7.869&8.264\\
\noalign{\smallskip}\hline\noalign{\smallskip}
\end{tabular}
\end{flushleft}
\end{table*}

\begin{table*}
\caption[]{Original Hipparcos $H_{\rm p}$ observations and their conversion to
$V$ band observations.}\label{phothpv}
\begin{flushleft}
\begin{tabular}{ccccccccc}
\hline\hline\noalign{\smallskip}
HJD&$H_{\rm p}$&rms&$V$&HJD&$H_{\rm p}$&rms&$V$\\
-2400000 &(mag)&(mag)&(mag)& &(mag)&(mag)&(mag)\\
\noalign{\smallskip}\hline\noalign{\smallskip}
47966.1374&7.2184&0.007&7.157&48393.5983&7.2257&0.008&7.164\\
47966.1517&7.2195&0.007&7.158&48404.6932&7.1996&0.012&7.138\\
47987.3742&7.1939&0.009&7.133&48404.7075&7.2290&0.008&7.168\\
47987.3885&7.2029&0.010&7.142&48404.9599&7.2330&0.014&7.172\\
47987.4630&7.1936&0.007&7.132&48404.9742&7.2168&0.008&7.156\\
47987.4774&7.2184&0.008&7.157&48405.0487&7.2312&0.014&7.170\\
48013.4952&7.1901&0.007&7.129&48569.9541&7.1610&0.006&7.100\\
48013.5095&7.1978&0.007&7.137&48569.9684&7.1657&0.013&7.104\\
48013.5840&7.2009&0.009&7.140&48570.0573&7.1574&0.011&7.096\\
48013.5984&7.1826&0.009&7.121&48586.1305&7.2339&0.010&7.173\\
48147.8756&7.2272&0.008&7.166&48586.1449&7.2120&0.008&7.151\\
48147.8900&7.2135&0.006&7.152&48586.2194&7.2232&0.009&7.162\\
48147.9646&7.2304&0.008&7.169&48586.4115&7.2242&0.007&7.163\\
48147.9789&7.2006&0.007&7.139&48705.7530&7.1657&0.008&7.104\\
48148.0534&7.2249&0.008&7.164&48705.7673&7.1590&0.008&7.098\\
48164.7628&7.2211&0.009&7.160&48705.8418&7.1182&0.013&7.057\\
48164.7771&7.1966&0.007&7.135&48731.2523&7.1546&0.006&7.093\\
48164.8517&7.1594&0.016&7.098&48731.2666&7.1673&0.006&7.106\\
48164.8660&7.2027&0.010&7.141&48753.2881&7.2043&0.007&7.143\\
48194.3522&7.1997&0.007&7.138&48753.3025&7.1939&0.006&7.133\\
48194.3665&7.1866&0.007&7.125&48753.3770&7.2045&0.009&7.143\\
48316.8239&7.2018&0.013&7.141&48753.3913&7.1985&0.008&7.137\\
48316.8382&7.1959&0.010&7.135&48935.2322&7.2130&0.008&7.152\\
48317.1794&7.1975&0.010&7.136&48935.2465&7.2100&0.008&7.149\\
48317.1937&7.1878&0.011&7.127&49047.3806&7.1871&0.012&7.126\\
48317.2683&7.1867&0.012&7.125&49047.3949&7.1602&0.009&7.099\\
48317.2826&7.1824&0.013&7.121&49047.6431&7.1776&0.010&7.116\\
48322.9710&7.2103&0.007&7.149&49047.6575&7.1718&0.009&7.111\\
48323.0455&7.1897&0.009&7.128&49047.7320&7.1710&0.009&7.110\\
48323.0598&7.2171&0.008&7.156&49047.7463&7.1712&0.010&7.110\\
48323.3122&7.2063&0.008&7.145&49047.8209&7.1732&0.009&7.112\\
48323.3265&7.2085&0.010&7.147&49047.8352&7.1607&0.008&7.099\\
48323.4011&7.1930&0.007&7.132&49053.4135&7.1785&0.008&7.117\\
48323.4154&7.1903&0.006&7.129&49053.4279&7.2007&0.007&7.139\\
48323.4900&7.2006&0.007&7.139&49053.5024&7.1748&0.009&7.114\\
48323.5043&7.2062&0.009&7.145&49053.5167&7.1760&0.011&7.115\\
48359.4706&7.2092&0.009&7.148&49053.5913&7.1690&0.008&7.108\\
48393.3316&7.2130&0.010&7.152&49053.6056&7.1613&0.010&7.100\\
48393.4062&7.2248&0.007&7.164&49053.8561&7.1615&0.007&7.100\\
48393.4205&7.2077&0.012&7.146&49053.8704&7.1701&0.007&7.109\\
48393.4951&7.2170&0.008&7.156&49053.9450&7.1604&0.011&7.099\\
48393.5094&7.2208&0.009&7.160&49054.0339&7.1688&0.009&7.108\\
48393.5840&7.2114&0.006&7.150& \\
\noalign{\smallskip}\hline\noalign{\smallskip}
\end{tabular}
\end{flushleft}
\end{table*}

\begin{table*}
\caption[]{Individual ASAS-3 $V$ band observations from diaphragm 1
without large errors.}\label{photv}
\begin{flushleft}
\begin{tabular}{ccccccccc}
\hline\hline\noalign{\smallskip}
HJD&$V$&rms&HJD&$V$&rms&HJD&$V$&rms\\
-2400000&(mag)&(mag)&-2400000&(mag)&(mag)&-2400000&(mag)&(mag)\\
\noalign{\smallskip}\hline\noalign{\smallskip}
52552.8776&7.156& 0.034& 52970.7859&7.118& 0.037& 54107.7196&7.126& 0.037\\
52557.8567&7.145& 0.036& 52973.8098&7.129& 0.032& 54127.7145&7.114& 0.037\\
52559.8729&7.148& 0.035& 52975.7732&7.089& 0.038& 54141.7008&7.144& 0.038\\
52577.8040&7.143& 0.036& 52979.7768&7.147& 0.033& 54146.6021&7.119& 0.038\\
52623.7460&7.175& 0.038& 52986.7882&7.143& 0.039& 54152.5591&7.095& 0.037\\
52625.7454&7.124& 0.036& 52989.7358&7.101& 0.037& 54154.5939&7.160& 0.037\\
52637.7183&7.145& 0.036& 52991.7752&7.155& 0.038& 54174.6296&7.149& 0.040\\
52641.7108&7.132& 0.036& 52996.7692&7.128& 0.036& 54188.6063&7.138& 0.039\\
52645.7047&7.106& 0.038& 53001.7538&7.084& 0.031& 54201.5468&7.113& 0.038\\
52650.6776&7.156& 0.036& 53006.6958&7.099& 0.034& 54204.5454&7.101& 0.039\\
52652.6885&7.104& 0.036& 53008.7208&7.097& 0.039& 54216.5315&7.094& 0.039\\
52660.6893&7.120& 0.035& 53016.7075&7.141& 0.039& 54233.4825&7.115& 0.037\\
52664.6776&7.115& 0.037& 53018.7030&7.140& 0.040& 54235.4863&7.100& 0.039\\
52666.6750&7.114& 0.040& 53028.7644&7.105& 0.040& 54247.4538&7.102& 0.039\\
52670.6739&7.121& 0.038& 53031.6263&7.132& 0.034& 54365.8936&7.105& 0.036\\
52674.6623&7.153& 0.036& 53035.6532&7.137& 0.040& 54383.8650&7.068& 0.038\\
52676.6485&7.148& 0.037& 53046.6032&7.082& 0.036& 54412.8547&7.141& 0.038\\
52678.6339&7.089& 0.039& 53049.6167&7.150& 0.038& 54436.7788&7.100& 0.039\\
52681.6881&7.163& 0.040& 53054.5926&7.099& 0.035& 54455.7297&7.104& 0.039\\
52691.6242&7.104& 0.039& 53058.6732&7.094& 0.038& 54470.6969&7.145& 0.039\\
52693.6365&7.140& 0.038& 53060.6989&7.103& 0.039& 54479.6571&7.085& 0.040\\
52697.6305&7.101& 0.036& 53068.5635&7.144& 0.040& 54492.7149&7.086& 0.039\\
52699.6208&7.132& 0.039& 53077.5445&7.122& 0.034& 54494.7342&7.095& 0.040\\
52701.6187&7.164& 0.036& 53102.5989&7.143& 0.040& 54537.5394&7.123& 0.040\\
52703.6144&7.108& 0.038& 53107.5350&7.144& 0.040& 54558.6078&7.088& 0.039\\
52707.6045&7.169& 0.037& 53109.5477&7.132& 0.039& 54570.4912&7.088& 0.040\\
52718.5269&7.109& 0.039& 53111.5476&7.090& 0.040& 54601.4685&7.134& 0.039\\
52734.5519&7.138& 0.039& 53113.5671&7.140& 0.040& 54762.8482&7.127& 0.037\\
52740.5074&7.158& 0.039& 53115.5672&7.140& 0.035& 54774.8284&7.138& 0.038\\
52751.5004&7.143& 0.040& 53123.4834&7.093& 0.040& 54778.8182&7.075& 0.038\\
52760.5200&7.134& 0.037& 53125.5273&7.120& 0.038& 54801.7777&7.130& 0.037\\
52786.4622&7.134& 0.038& 53133.5104&7.144& 0.038& 54804.7718&7.108& 0.037\\
52911.8901&7.093& 0.037& 53144.4748&7.096& 0.034& 54807.7563&7.145& 0.036\\
52916.8806&7.097& 0.037& 53291.8729&7.082& 0.040& 54816.7776&7.087& 0.040\\
52926.8593&7.153& 0.037& 53298.8666&7.085& 0.039& 54819.7738&7.145& 0.039\\
52932.8498&7.150& 0.034& 53333.8126&7.149& 0.040& 54822.7171&7.091& 0.039\\
52940.8298&7.147& 0.037& 53357.7288&7.114& 0.038& 54825.7243&7.129& 0.040\\
52943.8327&7.099& 0.039& 53459.5662&7.079& 0.038& 54831.7039&7.138& 0.037\\
52949.8286&7.095& 0.038& 53466.5452&7.079& 0.039& 54834.6914&7.111& 0.036\\
52952.8292&7.148& 0.040& 53469.5526&7.129& 0.038& 54836.7857&7.091& 0.038\\
52955.8312&7.084& 0.036& 53650.8382&7.162& 0.040                         \\
52963.8062&7.104& 0.037& 54091.7187&7.096& 0.040                         \\
\noalign{\smallskip}\hline\noalign{\smallskip}
\end{tabular}
\end{flushleft}
\end{table*}

\begin{table*}
\caption[]{Individual \ubv\ band observations.}\label{photubv}
\begin{flushleft}
\begin{tabular}{ccccccccc}
\hline\hline\noalign{\smallskip}
HJD&Weight&$V$&$B$&$U$&\bv&\ub&$X$&$\tria X$ \\
-2400000& &(mag)&(mag)&(mag)&(mag)&(mag))\\
\noalign{\smallskip}\hline\noalign{\smallskip}
Hvar\\
\noalign{\smallskip}\hline\noalign{\smallskip}
55574.4833&1.00& 7.158& 7.408& 7.424& 0.250& 0.016& 1.229& 0.044\\
55574.4901&1.00& 7.168& 7.415& 7.458& 0.247& 0.043& 1.235& 0.042\\
55574.4990&1.00& 7.160& 7.410& 7.432& 0.250& 0.022& 1.246& 0.039\\
55578.4494&1.00& 7.121& 7.399& 7.447& 0.278& 0.048& 1.225& 0.052\\
55578.4566&1.00& 7.125& 7.405& 7.434& 0.280& 0.029& 1.224& 0.049\\
55578.4621&1.00& 7.124& 7.397& 7.436& 0.273& 0.039& 1.225& 0.047\\
55585.4986&1.00& 7.098& 7.391& 7.432& 0.293& 0.041& 1.308& 0.031\\
55585.5069&1.00& 7.101& 7.371& 7.403& 0.270& 0.032& 1.335& 0.029\\
55585.5152&1.00& 7.106& 7.390& 7.435& 0.284& 0.045& 1.366& 0.027\\
55585.5165&1.00& 7.111& 7.393& 7.440& 0.282& 0.047& 1.371& 0.027\\
55586.4851&1.00& 7.124& 7.398& 7.424& 0.274& 0.026& 1.280& 0.034\\
55586.4948&1.00& 7.131& 7.399& 7.433& 0.268& 0.034& 1.306& 0.032\\
55586.5039&1.00& 7.130& 7.409& 7.432& 0.279& 0.023& 1.334& 0.029\\
55594.4271&1.50& 7.175& 7.417& 7.433& 0.242& 0.016& 1.229& 0.044\\
55594.4356&1.50& 7.182& 7.417& 7.430& 0.235& 0.013& 1.235& 0.041\\
55594.4444&1.50& 7.184& 7.426& 7.435& 0.242& 0.009& 1.246& 0.039\\
55596.4157&1.00& 7.160& 7.424& 7.441& 0.264& 0.017& 1.226& 0.046\\
55596.4243&1.00& 7.161& 7.426& 7.437& 0.265& 0.011& 1.230& 0.043\\
55596.4325&1.00& 7.154& 7.434& 7.433& 0.280&-0.001& 1.238& 0.041\\
55597.3054&1.50& 7.136& 7.399& 7.421& 0.263& 0.022& 1.468& 0.121\\
55597.3134&1.50& 7.133& 7.397& 7.440& 0.264& 0.043& 1.425& 0.111\\
55597.3212&1.50& 7.147& 7.401& 7.434& 0.254& 0.033& 1.389& 0.102\\
55599.3682&0.50& 7.121& 7.391& 7.425& 0.270& 0.034& 1.245& 0.063\\
55599.3724&0.50& 7.129& 7.400& 7.444& 0.271& 0.044& 1.239& 0.061\\
55599.3754&0.50& 7.125& 7.395& 7.431& 0.270& 0.036& 1.236& 0.059\\
55645.3479&0.50& 7.152& 7.406& 7.437& 0.254& 0.031& 1.365& 0.028\\
55645.3561&0.50& 7.156& 7.412& 7.436& 0.256& 0.024& 1.400& 0.026\\
55645.3604&0.50& 7.167& 7.418& 7.452& 0.251& 0.034& 1.421& 0.025\\
55652.3121&0.50& 7.147& 7.392& 7.409& 0.245& 0.017& 1.310& 0.031\\
55652.3220&0.50& 7.167& 7.428& 7.454& 0.261& 0.026& 1.342& 0.029\\
55652.3250&0.50& 7.174& 7.419& 7.449& 0.245& 0.030& 1.353& 0.028\\
55654.3169&1.00& 7.160& 7.409& 7.451& 0.249& 0.042& 1.344& 0.029\\
55654.3253&1.00& 7.163& 7.410& 7.434& 0.247& 0.024& 1.377& 0.027\\
55654.3333&1.00& 7.161& 7.411& 7.428& 0.250& 0.017& 1.413& 0.025\\
55655.3064&1.00& 7.136& 7.405& 7.440& 0.269& 0.035& 1.319& 0.031\\
55655.3149&1.00& 7.131& 7.404& 7.443& 0.273& 0.039& 1.347& 0.029\\
55655.3230&1.00& 7.123& 7.396& 7.430& 0.273& 0.034& 1.379& 0.027\\
55657.3373&1.00& 7.124& 7.401& 7.434& 0.277& 0.033& 1.479& 0.023\\
55657.3423&1.00& 7.128& 7.418& 7.480& 0.290& 0.062& 1.510& 0.021\\
55657.3452&1.00& 7.116& 7.415& 7.486& 0.299& 0.071& 1.529& 0.021\\
55658.3436&0.50& 7.173& 7.424& 7.458& 0.251& 0.034& 1.537& 0.021\\
55658.3477&0.50& 7.165& 7.419& 7.454& 0.254& 0.035& 1.566& 0.020\\
55658.3530&0.50& 7.167& 7.412& 7.433& 0.245& 0.021& 1.607& 0.019\\
55660.3295&1.00& 7.179& 7.420& 7.436& 0.241& 0.016& 1.483& 0.022\\
55660.3307&1.00& 7.185& 7.421& 7.437& 0.236& 0.016& 1.490& 0.022\\
55660.3349&1.00& 7.190& 7.421& 7.446& 0.231& 0.025& 1.517& 0.021\\
55661.3187&1.00& 7.165& 7.410& 7.439& 0.245& 0.029& 1.439& 0.024\\
55661.3273&1.00& 7.160& 7.410& 7.455& 0.250& 0.045& 1.487& 0.022\\
55661.3359&1.00& 7.154& 7.409& 7.454& 0.255& 0.045& 1.543& 0.020\\
55662.3278&0.50& 7.123& 7.378& 7.408& 0.255& 0.030& 1.507& 0.022\\
55662.3407&0.50& 7.108& 7.363& 7.395& 0.255& 0.032& 1.599& 0.019\\
55662.3420&0.50& 7.116& 7.371& 7.420& 0.255& 0.049& 1.609& 0.019\\
55848.5816&1.00& 7.145& 7.415& 7.480& 0.270& 0.065& 1.690& 0.175\\
55848.5922&1.00& 7.145& 7.422& 7.455& 0.277& 0.033& 1.599& 0.152\\
55848.5979&1.00& 7.118& 7.384& 7.394& 0.266& 0.010& 1.556& 0.142\\
55848.6654&1.00& 7.143& 7.416& 7.458& 0.273& 0.042& 1.269& 0.071\\
55850.5676&0.50& 7.097& 7.369& 7.406& 0.272& 0.037& 1.779& 0.198\\
55850.5774&0.50& 7.113& 7.385& 7.413& 0.272& 0.028& 1.680& 0.172\\
55850.5863&0.50& 7.119& 7.390& 7.440& 0.271& 0.050& 1.604& 0.154\\
55851.5815&1.00& 7.145& 7.408& 7.428& 0.263& 0.020& 1.622& 0.158\\
55851.5899&1.00& 7.149& 7.413& 7.442& 0.264& 0.029& 1.557& 0.142\\
55851.5980&1.00& 7.156& 7.413& 7.444& 0.257& 0.031& 1.502& 0.129\\
\noalign{\smallskip}\hline\noalign{\smallskip}
\end{tabular}
\end{flushleft}
\end{table*}

\setcounter{table}{5}
\begin{table*}
\caption[]{(cont.) Individual \ubv\ band observations.}\label{photubv}
\begin{flushleft}
\begin{tabular}{ccccccccc}
\hline\hline\noalign{\smallskip}
HJD&Weight&$V$&$B$&$U$&\bv&\ub&$X$&$\tria X$ \\
-2400000& &(mag)&(mag)&(mag)&(mag)&(mag))\\
\noalign{\smallskip}\hline\noalign{\smallskip}
55852.6273&1.50& 7.179& 7.429& 7.444& 0.250& 0.015& 1.347& 0.092\\
55852.6355&1.50& 7.175& 7.423& 7.443& 0.248& 0.020& 1.319& 0.085\\
55852.6445&1.50& 7.178& 7.423& 7.441& 0.245& 0.018& 1.294& 0.078\\
55853.6267&1.00& 7.184& 7.422& 7.426& 0.238& 0.004& 1.340& 0.090\\
55853.6346&1.00& 7.177& 7.422& 7.438& 0.245& 0.016& 1.314& 0.083\\
55853.6426&1.00& 7.185& 7.431& 7.442& 0.246& 0.011& 1.292& 0.077\\
55858.6678&1.50& 7.161& 7.415& 7.440& 0.254& 0.025& 1.230& 0.056\\
55858.6757&1.50& 7.155& 7.413& 7.437& 0.258& 0.024& 1.226& 0.053\\
55879.5943&1.50& 7.182& 7.424& 7.435& 0.242& 0.011& 1.250& 0.065\\
55879.6022&1.50& 7.183& 7.436& 7.444& 0.253& 0.008& 1.240& 0.061\\
55879.6098&1.50& 7.190& 7.423& 7.434& 0.233& 0.011& 1.232& 0.057\\
55881.6494&1.50& 7.121& 7.393& 7.423& 0.272& 0.030& 1.238& 0.041\\
55881.6545&1.50& 7.115& 7.394& 7.426& 0.279& 0.032& 1.245& 0.039\\
55881.6579&1.50& 7.118& 7.401& 7.432& 0.283& 0.031& 1.250& 0.038\\
55938.4426&1.50& 7.171& 7.420& 7.437& 0.249& 0.017& 1.242& 0.062\\
55938.4505&1.50& 7.175& 7.418& 7.430& 0.243& 0.012& 1.233& 0.058\\
55938.4552&1.50& 7.176& 7.420& 7.436& 0.244& 0.016& 1.229& 0.056\\
55939.3829&1.50& 7.133& 7.402& 7.430& 0.269& 0.028& 1.398& 0.104\\
55939.3882&1.50& 7.134& 7.408& 7.433& 0.274& 0.025& 1.375& 0.099\\
55939.3916&1.50& 7.136& 7.403& 7.430& 0.267& 0.027& 1.361& 0.095\\
55942.4525&1.50& 7.153& 7.409& 7.430& 0.256& 0.021& 1.225& 0.052\\
55942.4592&1.50& 7.160& 7.413& 7.436& 0.253& 0.023& 1.224& 0.050\\
55942.4674&1.50& 7.161& 7.411& 7.434& 0.250& 0.023& 1.225& 0.047\\
55943.4365&1.50& 7.177& 7.413& 7.427& 0.236& 0.014& 1.234& 0.058\\
55943.4449&1.50& 7.179& 7.420& 7.436& 0.241& 0.016& 1.227& 0.054\\
55943.4499&1.50& 7.178& 7.418& 7.431& 0.240& 0.013& 1.225& 0.052\\
55945.3602&1.50& 7.157& 7.409& 7.439& 0.252& 0.030& 1.429& 0.111\\
55945.3665&1.50& 7.164& 7.413& 7.437& 0.249& 0.024& 1.398& 0.104\\
55945.3698&1.50& 7.168& 7.415& 7.431& 0.247& 0.016& 1.384& 0.101\\
56001.3240&1.50& 7.177& 7.421& 7.430& 0.244& 0.009& 1.241& 0.040\\
56001.3316&1.50& 7.173& 7.415& 7.426& 0.242& 0.011& 1.251& 0.038\\
56001.3366&1.50& 7.175& 7.419& 7.430& 0.244& 0.011& 1.260& 0.037\\
56002.3518&1.50& 7.184& 7.418& 7.435& 0.234& 0.017& 1.301& 0.032\\
56002.3596&1.50& 7.176& 7.415& 7.434& 0.239& 0.019& 1.324& 0.030\\
56002.3643&1.50& 7.186& 7.435& 7.452& 0.249& 0.017& 1.340& 0.029\\
56013.3093&0.50& 7.142& 7.408& 7.436& 0.266& 0.028& 1.273& 0.035\\
56013.3142&0.50& 7.148& 7.409& 7.431& 0.261& 0.022& 1.284& 0.034\\
56013.3203&0.50& 7.146& 7.396& 7.424& 0.250& 0.028& 1.299& 0.032\\
56015.3123&0.50& 7.181& 7.432& 7.443& 0.251& 0.011& 1.293& 0.033\\
56015.3185&0.50& 7.184& 7.439& 7.475& 0.255& 0.036& 1.310& 0.031\\
56015.3245&0.50& 7.192& 7.435& 7.446& 0.243& 0.011& 1.329& 0.030\\
\noalign{\smallskip}\hline\noalign{\smallskip}
SAAO\\
\noalign{\smallskip}\hline\noalign{\smallskip}
55567.4069&1.00& 7.136& 7.417& 7.431& 0.281& 0.014& 1.453& 0.002\\
55567.4146&1.00& 7.139& 7.425& 7.429& 0.286& 0.004& 1.420& 0.002\\
55567.4217&1.00& 7.144& 7.420& 7.427& 0.276& 0.007& 1.394& 0.002\\
55570.5248&0.50& 7.175& 7.434& 7.425& 0.259&-0.009& 1.415& 0.004\\
55573.4850&1.00& 7.144& 7.409& 7.415& 0.265& 0.006& 1.331& 0.003\\
55573.4914&1.00& 7.122& 7.394& 7.426& 0.272& 0.032& 1.343& 0.003\\
55573.4981&1.00& 7.119& 7.400& 7.415& 0.281& 0.015& 1.357& 0.003\\
55574.5248&1.00& 7.180& 7.428& 7.462& 0.248& 0.034& 1.461& 0.004\\
55574.5308&1.00& 7.154& 7.409& 7.426& 0.255& 0.017& 1.491& 0.004\\
55574.5371&1.00& 7.166& 7.403& 7.421& 0.237& 0.018& 1.526& 0.005\\
55575.5287&1.00& 7.180& 7.421& 7.433& 0.241& 0.012& 1.494& 0.005\\
55575.5345&1.00& 7.169& 7.429& 7.415& 0.260&-0.014& 1.528& 0.005\\
55575.5408&1.00& 7.191& 7.440& 7.426& 0.249&-0.014& 1.568& 0.005\\
55577.5457&1.50& 7.135& 7.415& 7.412& 0.280&-0.003& 1.646& 0.006\\
55578.5443&0.50& 7.130& 7.401& 7.420& 0.271& 0.019& 1.657& 0.006\\
55578.5505&0.50& 7.133& 7.420& 7.436& 0.287& 0.016& 1.712& 0.006\\
55579.5226&1.00& 7.110& 7.407& 7.431& 0.297& 0.024& 1.522& 0.005\\
55579.5288&1.00& 7.125& 7.401& 7.418& 0.276& 0.017& 1.560& 0.005\\
55579.5345&1.00& 7.118& 7.382& 7.422& 0.264& 0.040& 1.601& 0.005\\
\noalign{\smallskip}\hline\noalign{\smallskip}
\end{tabular}
\end{flushleft}
\end{table*}

\end{appendix}
\end{document}